\documentclass[12pt,preprint]{aastex}

\def\lea{\mathrel{<\kern-1.0em\lower0.9ex\hbox{$\sim$}}}
\def\gea{\mathrel{>\kern-1.0em\lower0.9ex\hbox{$\sim$}}}

\slugcomment{Submitted to The Astrophysical Journal}

\shorttitle{ACS Globular Clusters: Sgr Distance}
\shortauthors{Siegel et al.}

\begin{document}

\title{The ACS Survey of Galactic Globular Clusters XI: The Three-Dimensional Orientation of the Sagittarius
Dwarf Spheroidal Galaxy and its Globular Clusters}

\author{Michael H. Siegel}
\affil{Department of Astronomy and Astrophysics, Pennsylvania State University, 525 Davey 
Laboratory, University Park, PA 16802 \email{siegel@astro.psu.edu}}

\author{Steven R. Majewski}
\affil{Dept. of Astronomy, University of Virginia, P.O. Box 400325, Charlottesville, VA 22904-4325
\email{srm4n@virginia.edu} }

\author{David R. Law\altaffilmark{1}}
\affil{Department of Physics and Astronomy, University of California, Los Angeles, CA 90095}
\email{drlaw@astro.ucla.edu}

\author{Ata Sarajedini}
\affil{Department of Astronomy, University of Florida, 211 Bryant Space Science Center, Gainesville, FL 32611
\email{ata@astro.ufl.edu} }

\author{Aaron Dotter}
\affil{Department of Physics and Astronomy, Elliott Building, 3800 Finnerty Road, Victoria, BC V8P 5C2 Canada}
\email{dotter@uvic.ca}

\author{A. Mar\' \i n-Franch}
\affil{Instituto de Astrof\'\i sica de Canarias, V\'\i a L\'actea s/n, E-38200 La Laguna, Spain}
\email{amarin@iac.es}

\author{Brian Chaboyer}
\affil{Department of Physics and Astronomy, Dartmouth College, 6127 Wilder Laboratory, Hanover, NH 03755 
\email{chaboyer@heather.dartmouth.edu} }


\author{Jay Anderson}
\affil{Space Telescope Science Institute, 3700 San Martin Drive, Baltimore MD 21218
\email{jayander@stsci.edu} }

\author{Antonio Aparicio}
\affil{University of La Laguna and Instituto de Astrof\'\i sica de Canarias, E-38200 La Laguna, Spain
\email{antapaj@iac.es} }

\author{Luigi R. Bedin}
\affil{Space Telescope Science Institute, 3700 San Martin Drive, Baltimore MD 21218
\email{bedin@stsci.edu}}

\author{Maren Hempel}
\affil{Departamento de Astronomia y Astrofisica, Vic. Mackenna 4860 Casilla 306, Santiago, 22, Chile
\email{mhempel@astro.puc.cl}}


\author{Antonino Milone}
\affil{Dipartimento di Astronomia, Universit\`{a} di Padova, I-35122 Padova, Italy 
\email{antonino.milone@oapd.inaf.it}}

\author{Nathaniel Paust}
\affil{Space Telescope Science Institute, 3700 San Martin Drive, Baltimore MD 21218
\email{paust@stsci.edu} }

\author{Giampaolo Piotto}
\affil{Dipartimento di Astronomia, Universit\`{a} di Padova, I-35122 Padova, Italy 
\email{giampaolo.piotto@unipd.it} }

\author{I. Neill Reid}
\affil{Space Telescope Science Institute, 3700 San Martin Drive, Baltimore MD 21218
\email{inr@stsci.edu} }

\author{Alfred Rosenberg}
\affil{Instituto de Astrof\'\i sica de Canarias, V\'\i a L\'{a}ctea s/n, E-38200 La Laguna, Spain
\email{alf@iac.es} }

\altaffiltext{1}{Hubble Fellow}

\begin{abstract}
We use observations from the {\sl Hubble Space Telescope} {\sl Advanced Camera for Surveys} 
({\it HST}/ACS)
study of Galactic globular clusters
to investigate the spatial 
distribution of the inner regions of 
the disrupting Sagittarius dwarf spheroidal galaxy (Sgr).  We combine previously published analyses of four Sgr member
clusters located near or in the Sgr core (M54, Arp 2, Terzan 7 and Terzan 8) with a new analysis of diffuse Sgr material
identified in the background of five low-latitude Galactic bulge clusters (NGC~6624, 6637, 6652, 6681 and 6809) observed as part
of the ACS survey.
By comparing the bulge cluster CMDs to our previous analysis of the M54/Sgr core, we estimate distances to these background features.
The combined data
from four Sgr member clusters and five Sgr background features provides nine
independent measures of the Sgr distance and, as a group, provide uniformly
measured and calibrated probes of different parts of the inner regions of Sgr
spanning
twenty degrees over the face of the disrupting dwarf.  This allows us, for the first time, to constrain the three
dimensional orientation of Sgr's disrupting core and globular cluster system and compare that orientation to the predictions of an $N$-body model of
tidal disruption.  The density and distance of Sgr debris is consistent with models that
favor a relatively high Sgr core mass and a slightly greater distance (28-30 kpc, with a mean of 29.4 kpc).  Our analysis also suggests that M54 is in the foreground
of Sgr by $\sim2$ kpc, projected on the center of the Sgr dSph. While this would imply a remarkable alignment of the cluster and the Sgr
nucleus along the line of sight, we can
not identify any systematic effect in our analysis that would falsely create the measured 2 kpc separation.
Finally, we find that the cluster Terzan 7 has the most discrepant distance (25 kpc)
among the four Sgr core clusters, which may suggest a different dynamical
history than the other Sgr core clusters.
\end{abstract}

\keywords{globular clusters: individual (M54; Terzan 7; Terzan 8; Arp 2; NGC~6624; NGC~6637; NGC~6652; NGC~6681;
NGC~6809); galaxies: individual (Sagittarius); galaxies: star clusters;
galaxies: stellar content}

\section{Introduction}
\label{s:intro}

The Sagittarius dwarf spheroidal galaxy (Sgr) is among the nearest known dwarf galaxies to the Milky Way.  Soon after its discovery (Ibata et al. 1994),
it became obvious that Sgr had an extended surface brightness profile consistent with the expected profile of a dwarf galaxy merging with the Milky
Way.  Its tidal tails have now been traced around the entire celestial sphere (see, e.g., Majewski et al. 2003, hereafter M03; Belokurov et al. 2006,
Yanny et al. 2009 and references therein; Correnti et al. 2010) and
it has become a paradigm of the hierarchical formation of the outer Milky Way.

Recent wide-field surveys and $N$-body merging simulations of Sgr
have done an excellent job of measuring and constraining the overall placement and dynamics of the extended tidal tails
of Sgr, with concomitant
insight into the shape of the Milky Way's dark matter halo (see, e.g., Ibata et al. 2001; Helmi 2004; Law et al. 2005, 2009; 
Johnston et al. 2005; Fellhauer et al. 2006;
Martinez-Delgado et al. 2007; Law \& Majewski, 2010a, hereafter LM10a),
the association of stellar structures within the halo (Bellazzini et al. 2003a,b; Law et al. 2005; LM10a; Law \& Majewski 2010b, 
hereafter LM10b)
and the status of potential Sgr debris within a few kpc of the Sun (Law et al. 2005; Yanny et al. 2009; LM10a).
However, many questions remain unanswered on Sgr's structure and chemodynamical history.
Sgr is the most elliptical of the dSphs and the origin of this ellipticity is unclear.
What is the relative distribution 
of dark and luminous matter?  Is the current Sgr core oblate or prolate?  Does the postulated bifurcation of the stream result from substructure in the satellite, as
suggested by LM10a, or from a disk structure in the initial satellite, as suggested by Penarrubia et al. (2010; but cf. Lokas
et al. 2010, Penarrubia et al. 2011, Frinchaboy et al. 2011, in prep)?
Are the globular clusters associated with Sgr (M54, Terzan 7, Terzan 8, Arp 2) bound or unbound to Sgr? Was the Sgr progenitor
similar to the SMC or LMC, as suggested by its stellar chemistry (Chou et al.
2010) and current total luminosity (Niederste-Ostholt et al. 2010)?  What is the mass of the bound remnant?  Is M54 the core of Sgr
or simply aligned with the core by chance? Or has it sunk to the core from dynamical friction?

Indeed, even the {\it distance} to Sagittarius is somewhat uncertain, with most estimates clumping close to the initial 24 kpc distance
reported by Ibata et al. (1994) but ranging up to 26 kpc (Monaco et al. 2004), 27 kpc (Layden \& Sarajedini 2000; Bellazzini et al. 2006; 
Kunder \& Chaboyer 2009; Sollima et al. 2010 ) or even 28 kpc (Siegel et al. 2007, hereafter Paper IV).

Insight into some or perhaps all of these issues could be obtained by studying the three-dimensional orientation of the
Sgr core and its incipient debris field.  However, study of the Sgr core and its emerging tidal debris arms is hampered by their location behind
a thick veil of foreground Milky Way bulge stars.  Large photometric surveys (e.g., Ibata et al. 1994; M03)
have shown the core to be elongated.  However, these studies have not been able to resolve definitively the distance to the Sgr
core, its three-dimensional shape or the status of M54 as the core of Sgr or a chance alignment.

In this paper, we use data 
from the ACS Survey of Galactic Globular Clusters (Sarajedini et al. 2007, hereafter Paper I)
to overcome some of these difficulties and provide
new insight into the spatial distribution of Sgr's inner regions.
The high spatial resolution
provided by the {\it Hubble Space Telescope} (HST) can overcome the crowding issues while the high precision of the deep ACS photometry can delineate faint
photometric sequences that are undetectable from the ground.  This was demonstrated by our study of the stellar populations in the Sgr core (Paper IV),
which easily identified previously unseen features in the Sgr CMD.

Using the unique and powerful ACS dataset, we now move beyond the Sgr center to examine the larger structure of the dSph.
In addition to four 
classical
Sgr member clusters that are close to or in the Sgr core (M54, Terzan 7, Terzan 8, Arp 2; Da Costa \& Armandroff 1995), 
we have identified Sgr debris in the background
of five Milky Way bulge clusters (\S\ref{s:background}).  The fortuitous combination of a well-measured foreground
cluster, a well-measured background population and a previously determined age-metallicity relationship (AMR) for Sgr (Paper IV) allows us to use
these background features to 
measure precise relative distances for five additional lines of site.  In combination with the member clusters, this allows us
to probe the shape of the bound core and emerging Sgr tidal debris over a large solid angle.  We compare the distances of Sgr
member clusters, the distance of Sgr features in the background of Milky Way bulge clusters
and the density of the background Sgr material 
to the $N$-body model of tidal disruption from LM10a to gain new
insight into the three-dimensional
shape of Sgr and its globular cluster system, as well as what their properties reveal about the
intrinsic shape, distance and dynamical evolution of the dSph
(\S\ref{s:3dsgr}).

\section{Observations and Data Reduction}
\label{s:obsred}

The ACS survey of Galactic Globular Clusters (Paper I) is a photometric study of 65
nearby globular clusters using the {\it Wide Field Channel} (WFC) of the {\it Advanced Camera for Surveys} (ACS) aboard
the HST. The objective is to use the high-precision color-magnitude diagrams (CMDs) provided by the {\sl HST/ACS/WFC}
to explore a host of issues in stellar evolution, cluster evolution, dynamics and the formation of the Milky Way.
To date, this program has provided detailed analyses of the CMDs of the NGC~1851, NGC~6366 and M54 globular clusters
(Milone et al. 2008; Paust et al. 2009; Paper IV).  We have also used the CMDs
for global analyses of relative cluster ages (Marin-Franch et al. 2009, hereafter Paper VII), luminosity
functions (Paust et al. 2010), horizontal branch morphologies (Dotter et al. 2010, hereafter Paper IX) and
cluster centers (Goldsbury et al. 2010).

We observed our target clusters
in the $F606W$ ($\sim$$V$) and $F814W$ ($\sim$$I$) filters
with {\sl HST/ACS/WFC} (Paper I).
Photometry was generated through new PSF methods (Anderson et al., 2008, hereafter Paper V)
and Vega-calibrated using
the charge-transfer efficiency corrections of Riess \& Mack (2004), calibration
procedures in Bedin et al. (2005) and zero points of Sirianni et al. (2005) with zero point
corrections updated to those of Bohlin et al. (2007).  The photometry
of isolated saturated stars on short exposures was
salvaged by summing all associated charge --- a procedure 
previously applied by Gilliland (2004).
For our clusters, the data pipeline provides twelve magnitudes of precise
photometry from nearly the
tip of the red giant branch (RGB) to several magnitudes below the main sequence turnoff (MSTO).
Uncertainties are approximately 1\% at $V\sim22$ and 10\% at $V\sim25$. 
We have cleaned the photometric catalogs of non-stellar objects and poorly measured stars using trends of
quality-of-fit against magnitude to select sources with the most star-like profiles and 
$<10$\% of the flux in their PSF aperture from other stars (see Paper V).  The photometric data are available
to the public from our archive.\footnote{\url{http://www.astro.ufl.edu/\textasciitilde ata/public\_hstgc/databases.html}}

\section{Distance Measures In the Sgr Field}
\label{s:distances}

\subsection{Sagittarius Member Clusters}
\label{ss:members}

Table \ref{t:members} lists the four Galactic globular clusters that are
within a few core radii of Sgr (given in M03 as $\sim3.8^{\circ}$).  
For each cluster, we
list coordinates, in degrees, in the Sgr coordinate
system ($\Lambda_{\odot}, B_{\odot}$) defined by the Euler angles given in Table 2 of M03, in which
$\Lambda_{\odot}$ is
defined as the angular distance along Sgr's orbit with positive $\Lambda_{\odot}$ in the direction of the trailing debris.
$B_{\odot}$ is the lateral angle from the Sgr meridian.  The Sgr
coordinate system allows a more natural analysis of the Sgr features relative to the dSph itself than other celestial coordinate
systems.

A number of other Galactic globular clusters have been suggested as members of Sagittarius (see, e.g., Dinescu et al. 2000; Palma et al. 2002; 
Bellazzini et al. 2002, 2003a,b; Cohen 2004; Carraro et al. 2007; Carraro 2009).  LM10b performed
a robust analysis and concluded that, besides the classical members, Whiting 1 and NGC~5634 have a high
probably of being part of Sgr and NGC~5053 and Pal 12 are moderately likely to be members.  Pal 12 and NGC~5053
were included in the ACS survey.  However, these clusters are situated 39 and 266 degrees along the Sgr tidal stream (LM10b), respectively.
While their distances could
provide insight into the shape of Sgr's tidal arms, we have left them out of our analysis, which is focused on the
three-dimensional orientation of the Sgr {\it core}.

\begin{center}
\begin{figure}[ht!]
\includegraphics[angle=0,scale=0.8]{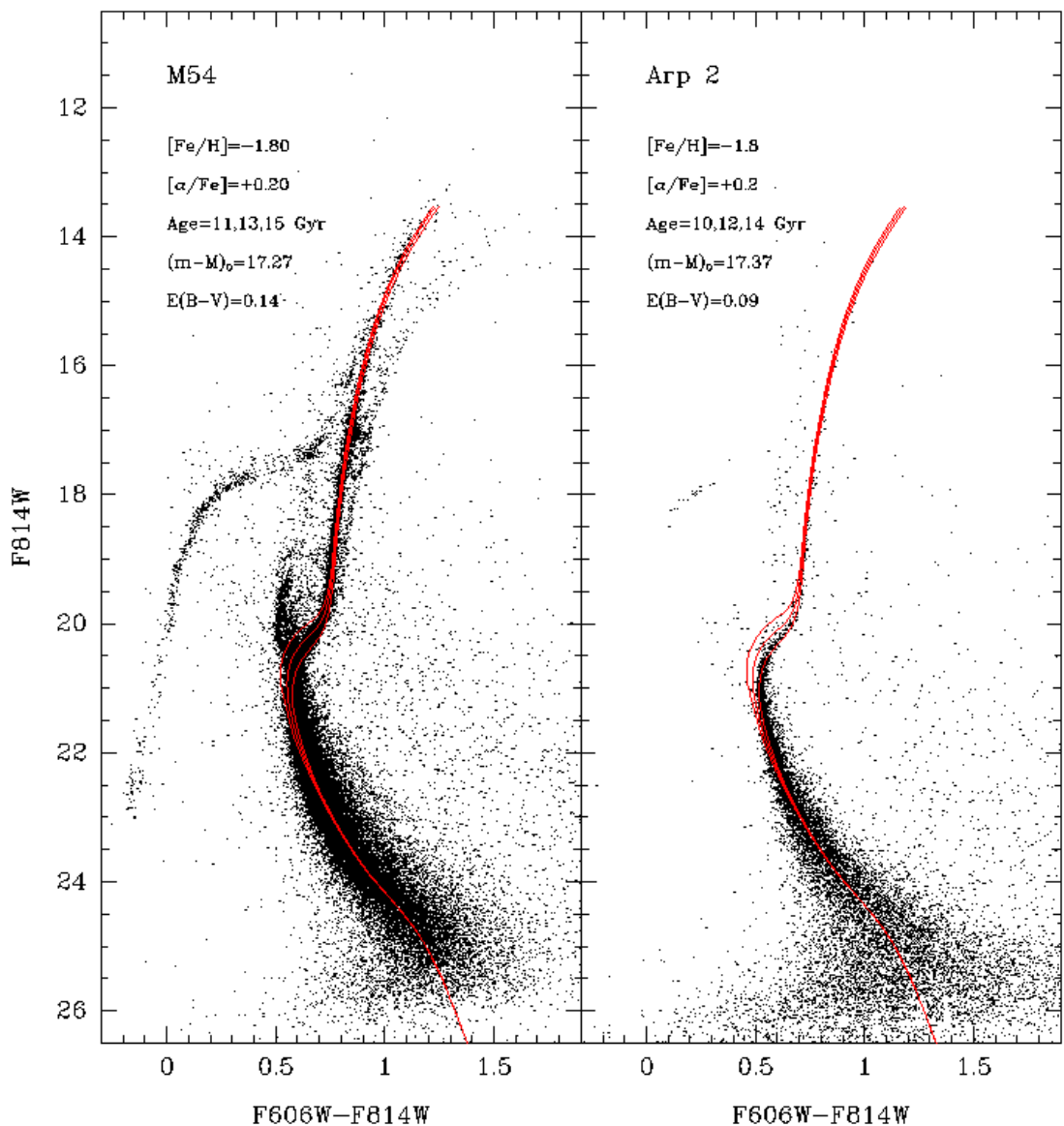}
\figcaption[f1.eps]{Isochrone fits to the clusters M54 and Arp 2 as derived in Paper IV and Paper IX, respectively.  The isochrones and
fits were adjusted slightly from Papers IV and IX to reflect recent spectroscopic abundance measures.\label{f:sgra}}
\end{figure}
\end{center}

\begin{center}
\begin{figure}[ht!]
\includegraphics[angle=0,scale=0.8]{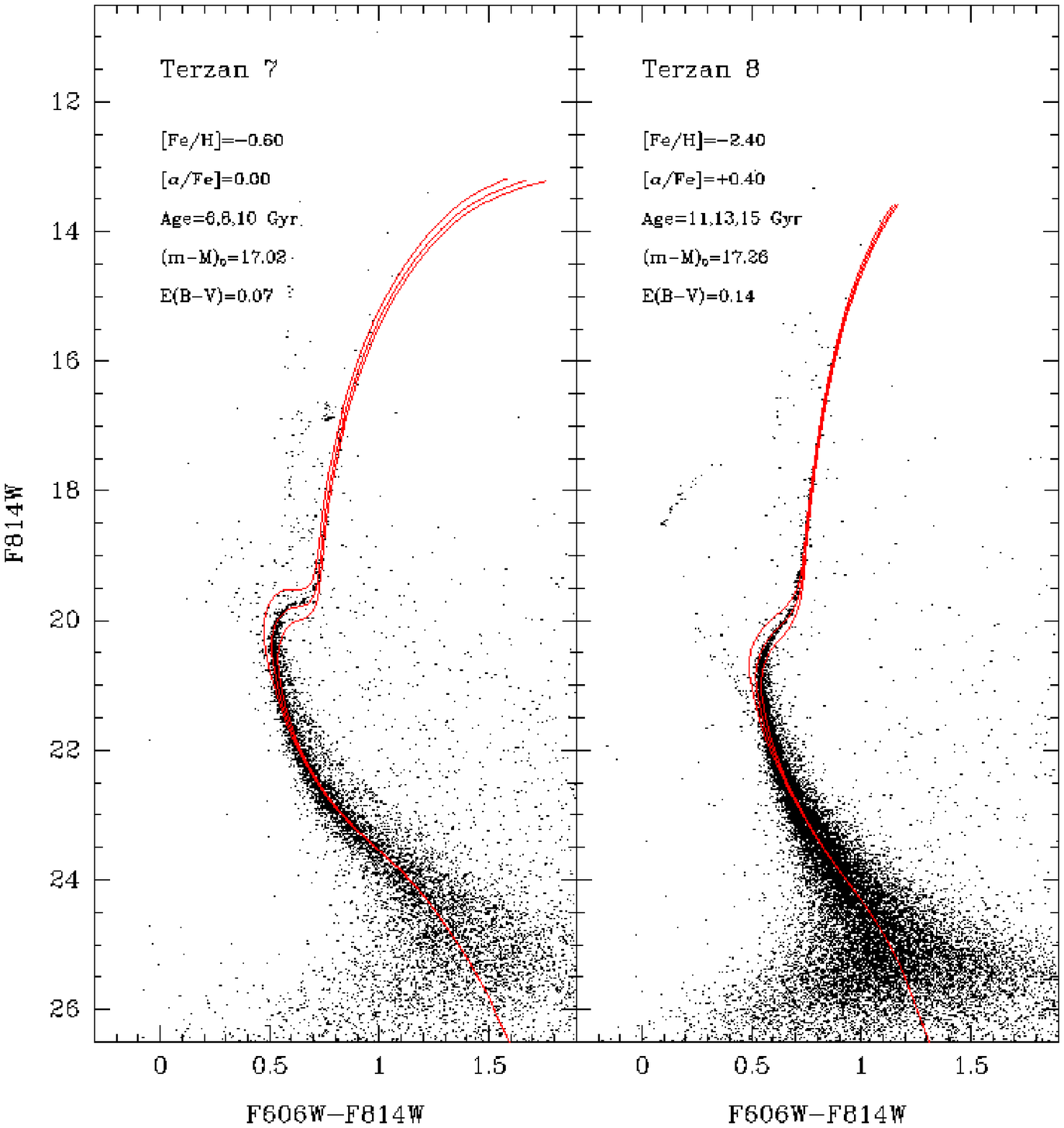}
\figcaption[f2.eps]{Same as Figure \ref{f:sgra} for Terzan 7 and Terzan 8.\label{f:sgrb}}
\end{figure}
\end{center}

Figures \ref{f:sgra} and \ref{f:sgrb} show the CMDs for the classical Sgr clusters.  For Terzan 7, Terzan 7 and Arp 2, we began our analysis with
the isochrones fitted to the cluster sequences
in Paper IX.  These isochrones, described in detail in Dotter et al. (2007), were fitted in Paper IX based on initial metallicity estimates
from the most recent catalog of Harris (1996, 2010 edition), tweaked slightly to better match the isochrones to the observed photometry.
For this paper, we have adjusted the fits slightly to reflect the most recent 
[Fe/H] and [$\alpha$/Fe] values measured for the Sgr clusters 
(see, e.g., Sbordone et al. 2005; Mottini et al. 2008), as listed in Table \ref{t:members}.
In all cases, the distances are based on the model-predicted absolute
magnitudes, scaled to an assumed value of the solar luminosity (3.4818 $\times$ 10$^{33}$ erg s$^{-1}$). Extinction values were determined directly from the isochrone
fits and converted into $E(B-V)$ values from the HST colors using the coefficients of Paper I. Distance uncertainties are difficult
to estimate because they depend upon both the mass-metallicity-luminosity relationship of the stars and the bolometric corrections.  We estimate
that the absolute distance uncertainties are approximately 0.1 mag while the relative distance uncertainties, which are the more
important to our differential analysis, are 0.05 mag or lower.

\begin{center}
\begin{deluxetable}{lcccccccc}
\tablewidth{0 pt}
\tablecaption{Sagittarius Member Clusters\label{t:members}}
\tablehead{
\colhead{Cluster} &
\colhead{$\Lambda_{\odot}$} &
\colhead{$B_{\odot}$} &
\colhead{[Fe/H]} &
\colhead{[$\alpha$/Fe]} &
\colhead{Age} &
\colhead{$(m-M)_{0}$} &
\colhead{$E(B-V)$} &
\colhead{Distance}\\
\colhead{} &
\colhead{($^{\circ}$)} &
\colhead{($^{\circ}$)} &
\colhead{} &
\colhead{} &
\colhead{} &
\colhead{} &
\colhead{} &
\colhead{(kpc)}
}
\startdata
\hline
Arp 2      & 7.2  & 0.4 & -1.89 & +0.31 & 13.5 & 17.37 & 0.09 & 29.4$\pm$0.7\\ 
Terzan 7   & 5.4  & 4.9 & -0.60 & -0.03 & 8.0  & 17.05 & 0.06 & 25.3$\pm$0.6\\ 
Terzan 8   & 10.3 & 3.8 & -2.30 & +0.37 & 13.0 & 17.26 & 0.14 & 28.3$\pm$0.7\\ 
M54        & 0.0  & 1.5 & -1.80 & +0.20 & 13.5 & 17.27 & 0.14 & 28.4$\pm$0.7\\ %
\hline
\enddata
\end{deluxetable}
\end{center}

Paper IX did not analyze M54 due to the multiple populations present in the field.  Paper IV derived a distance/reddening based
on the dominant metal-poor population of M54 itself.  As noted in \S\ref{s:m54stat}, there are some differences between the semi-empirical
isochrones used in Paper IV and the synthetic isochrones used in Paper IX and this paper.  However, these differences mostly affect the youngest
Sgr populations.  Using the most recent  synthetic isochrones, we find that the distance/age/abundance/reddening used in
Paper IV still provide an excellent fit to the photometry of the dominant metal-poor main sequence (see Figure \ref{f:sgra}).

\begin{center}
\begin{figure}[ht!]
\includegraphics[]{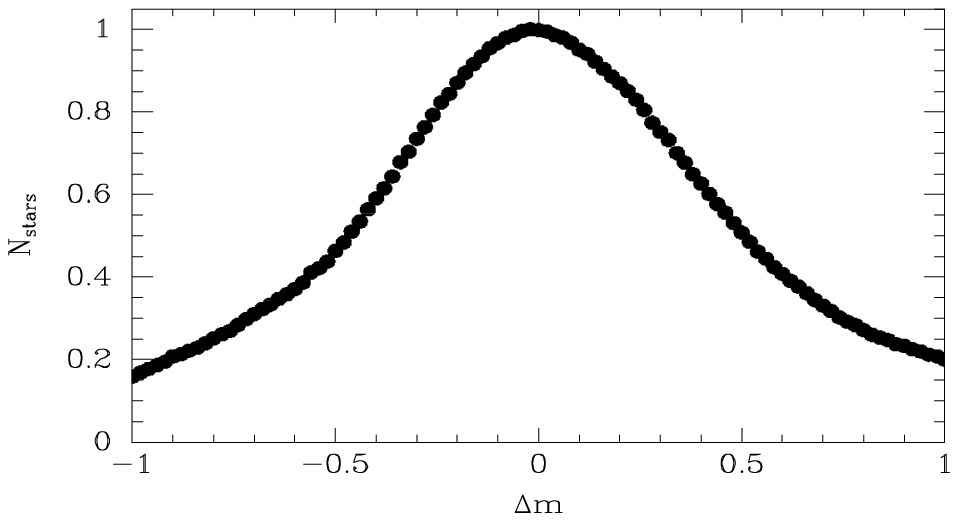}
\figcaption[f3.eps]{The normalized number of stars within 0.05 magnitudes of the Paper VII fiducial as a function of a magnitude
shift applied to the fiducial.  This shows the test applied to Terzan 7; the results for Arp 2 and Terzan 8 are similar.\label{f:magshift}}
\end{figure}
\end{center}

In an effort to detect any diffuse background Sgr tidal debris influencing our analysis of the clusters near the center of
Sgr, we took the cluster fiducials for Arp 2, Terzan 7 and Terzan 8
from Paper VII and shifted them in magnitude,
counting the number of stars within 0.05 magnitudes of the fiducial.  For a simple stellar population, like that expected in a typical
globular cluster, we would expect the count of stars 
to follow a Gaussian pattern, albeit with a slight increase at faint magnitudes from increasing photometric scatter near the fiducial and a 
tail at bright magnitudes due to unresolved binary stars.  If a significant second
population at one distance -- in this case Sgr -- were present, it might show up as a second bump or an asymmetry in the distribution, even if the
fiducial does not precisely
match the background population.  Detecting the Sgr background feature would allow us to constrain any effect it has upon the
isochrone fitting and measure any distance discrepancy between the Sgr 
member clusters and the stellar tidal debris stream.

For Arp 2, Terzan 7 and Terzan 8, the number counts follow a Gaussian distribution with no deviation larger than
the Poisson noise (see Figure \ref{f:magshift}).
This indicates either that (1) the clusters are in or near the mean distance of the Sgr stream; (2) that the Sgr background population is
simply too diffuse and/or (3) too close in color-magnitude space to the primary cluster sequence to show up distinctly in our analysis.  Given that
the distance measures in Table \ref{t:members} suggest a significant distance discrepancy for at least Terzan 7, (2) is the more
likely explanation.

\subsection{Sagittarius Background Features}
\label{s:background}

\begin{center}
\begin{figure}[ht!]
\plotone{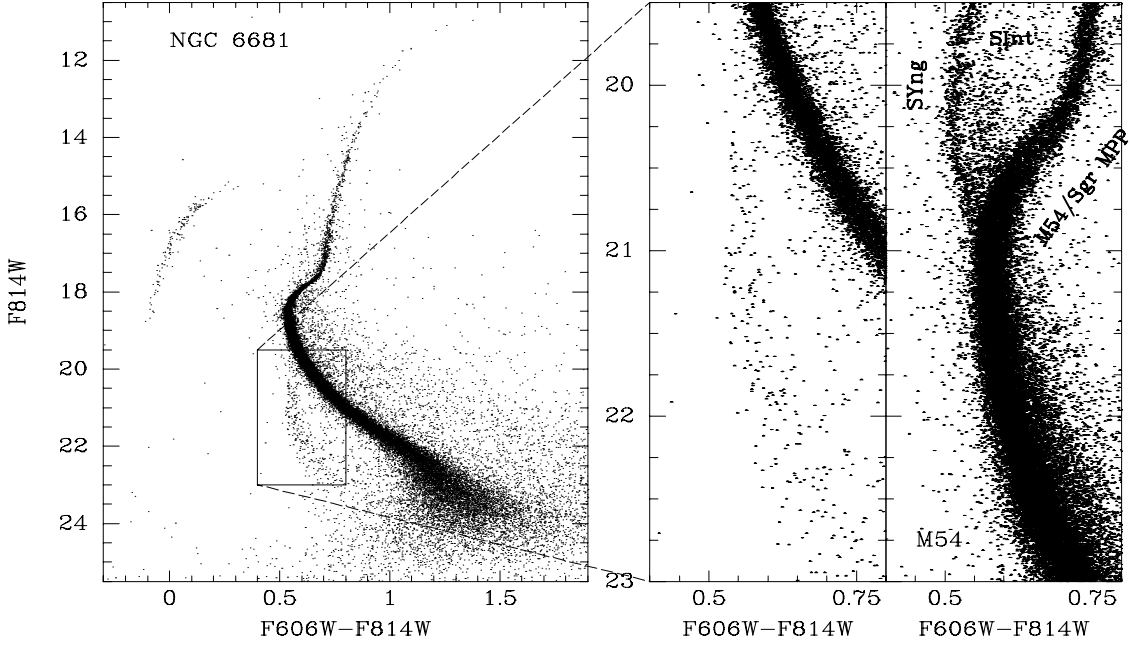}
\figcaption[f4.eps]{The background Sgr feature, as discovered in the field of NGC~6681.  The 
middle panel shows a zoomed-in view of
the second MS in the background of the bulge cluster.  The right panel shows a similar region of the CMD of M54 -- which either is
at the same distance or slightly in the foreground of the Sgr core -- shifted 0.15 mag in accordance with the distance measured in Table \ref{t:bgtab}.  We have labeled the young
(SYng) and intermediate (SInt) Sgr populations in M54's core, as defined in Paper IV and in the text.  The metal-poor population, which creates the dominant MSTO/SGB,
is a mix of M54's metal-poor population and Sgr's and is labeled appropriately (M54/Sgr MPP).
Note the similarity between the background
population of NGC~6681 and the intermediate-age populations of Sgr (which tend to be on the redward side of the broad M54/Sgr main sequence). Note also that the 
SYng population -- the brightest
and bluest Sgr feature -- is {\it not} seen in the Sgr background feature, indicating that this population is confined to the Sgr core. The breadth
of the Sgr feature in the NGC~6681 field is not the product of observational uncertainty but reflects genuine dispersion in the Sgr stellar populations.
\label{f:sgrbgdem}}
\end{figure}
\end{center}

In the course of examining the CMDs of the ACS survey globular clusters, we identified five bulge cluster CMDs that
appear to contain a faint secondary stellar sequence parallel to and below the cluster main sequence (Figure \ref{f:sgrbgdem}).
This secondary sequence is too bright and red to represent the white dwarf sequence of the cluster and has the appearance
of a diffuse main sequence.  It is broader than the photometric uncertainties would indicate for a simple stellar population,
which hints at a composite population.  Given that the suspect clusters lie in the foreground of the body and tidal stream of Sgr, it is likely
that this feature represents the diffuse Sgr tidal stream in the background of these clusters.
This Sgr background feature has been hinted at before (see, e.g., Mateo et al. 1996) but has not
been depicted as cleanly as shown in Figure \ref{f:sgrbgdem}.  The deep and precise ACS data delineate three magnitudes
of this secondary sequence.
This feature appears in all of the ACS survey clusters that are relatively close (within 11 degrees) to the Sgr core.

The fortuitous alignment of the Sgr background features with simple and well-measured
foreground clusters -- in combination with the consistent and precise photometry produced by our pipeline -- provides an unprecedented
opportunity to measure the distance of the Sgr core and/or tidal stream along multiple lines of sight across the face of the dSph.
While the foreground clusters generally have different abundances than typical Sgr core stars, the stars of the foreground MSs 
provide a useful
first estimate of the relative distances to the suspected Sgr features.  Using the fiducials from Paper VII,  
we measured by eye the color-magnitude shifts needed to move each foreground cluster's fiducial to overlap the suspected Sgr population.  These
relative measures confirm that the secondary sequences are at a distance consistent with that of the Sgr core.

\begin{center}
\begin{figure}[ht!]
\plotone{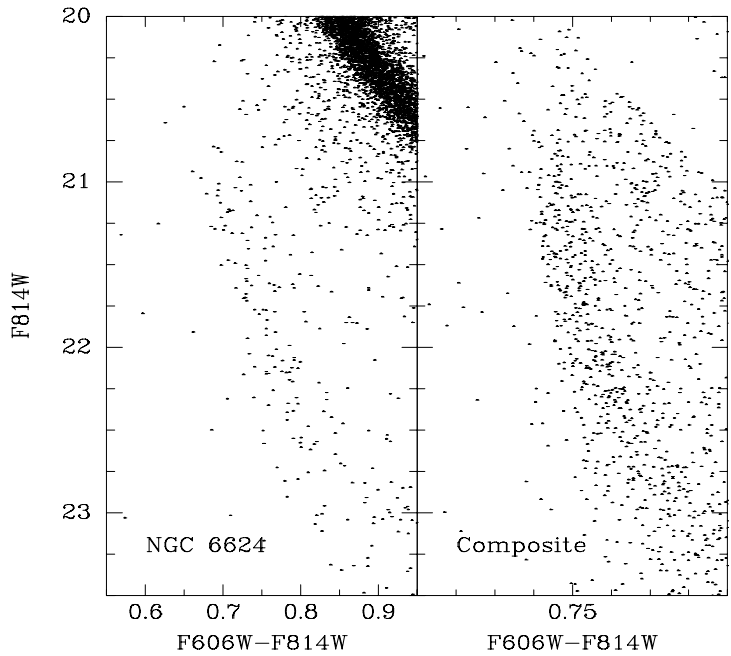}
\figcaption[f5.eps]{The enhancement of the Sgr background feature.  The left panel shows the CMD for NGC~6624.  The right shows
the composite CMD of all five background features, shifted to the same distance and reddening as the NGC~6624 background feature.\label{f:enhance}}
\end{figure}
\end{center}

We then refined these rough estimates by measuring the relative distance of each of the background features to each other.
We first removed from each color-magnitude
diagram those stars that lay within 0.1 magnitudes of the foreground clusters' fiducial, as defined in Paper VII.  We then
established NGC~6681, which has the strongest Sgr feature, as a template and overlayed the color-magnitude diagram of NGC~6624, which has
the second-strongest.  We shifted the
CMD of NGC~6624
in color-magnitude space by eye until the Sgr feature of NGC~6624 overlapped that of NGC~6681.  The data from the two clusters
were then combined and the cluster with the next-strongest feature overlayed and shifted.  This process continued through all five of the
bulge clusters with clean secondary sequences
until color-magnitude shifts were established for the entire sample.  This CMD shifting and adding enhances the Sgr feature (see Figure
\ref{f:enhance}) and allows to re-shift each cluster iteratively in the combined CMD to further refine the relative distance
estimates.

These relative measures were then calibrated to absolute distance by 
overlaying the isochrones fitted to the Sgr population from Paper IV.  The M54/Sgr field is a complex medley of populations.
Paper IV describes the Sgr field as a composite of seven populations (see their
Figure 2), the most prominent of which are shown in Figure \ref{f:sgrbgdem}:

$\bullet$ SVYng: a diffuse very young (100-800 Myr) population of super-solar abundance.

$\bullet$ SYng: a young (2.3 Gyr) population of solar abundance. This population could be of slightly younger age (1.75 Gyr)
if it has an enhanced He abundance
similar to the young population identified in $\omega$ Centauri (Carretta et al. 2010b).

$\bullet$ SInt: three intermediate age (4-7 Gyr) populations of intermediate abundance ([Fe/H]$ = -0.3$ to $-0.7$).

$\bullet$ Sgr MPP: an old (11 Gyr) metal-poor ([Fe/H]$ = -1.3$) population that significantly overlaps M54's metal-poor population.

$\bullet$ M54 MPP: the old metal-poor population of M54 itself (13 Gyr, [Fe/H]$= -1.8$), which significantly overlaps the Sgr MPP.

The Paper IV analysis has since been complemented by the spectroscopic survey of Carretta et al. (2010b), whose
spectroscopic metallicity distribution is similar to that described above (see their Figure 1).  However, they describe the M54 MPP population as having a slightly
higher metallicity of
[Fe/H]$= -1.55$ with a significant (0.19 dex) dispersion.
Frinchaboy et al. (in prep.) explore
metallicity distribution along Sgr's major and minor axes and shows clear evidence of a spatial trend in abundance (about 0.2-0.3 dex out to the radii of our
bulge cluster fields).  This trend could be critical to analysis of the Sgr background features.

The derived absolute distance of the Sgr CMD features, as measured by comparison to the Paper IV isochrones, is sensitive to which of these seven populations
are included in the analysis.
If we include all seven populations, the inferred distances would be many kpc farther than the distances measured for the
Sgr core clusters.  However, including all seven populations is likely inappropriate for the background features, which all lie outside of the
core of Sgr (defined as 3.7-3.9 degrees in M03) and are well way from the central M54 pointing.  No study, from either the ground or HST, has found
the youngest populations seen in the Paper IV analysis beyond these innermost regions of Sgr; and none of the five background
features has the bright blue SYng population seen in the M54 field (the brightest turnoff in Figure 1).

The aforementioned spectroscopic studies of Sgr's inner regions (Carretta et al. 2010b; Frinchaboy et al., in prep)
would favor using the more metal-poor of Sgr's seven populations.
The Frinchaboy et al. study, in particular, traces out the major axis of Sgr into regions overlapping our survey.  The metallicity distribution
in their outer regions would be best matched by a combination of 
the Sgr MPP population with the more metal poor SInt populations (those with [Fe/H] of $-0.5$ and $-0.7$).  Such a choice of isochrones
would also be consistent with the approximately 0.2 dex drop in metallicity between the core and the tidal arms indicated
by studies of the tidal arms (LM10a; Alard et al. 2001; Chou et al. 2007, 2010; Giuffrida et al. 2010).  We therefore have
chosen these three populations -- Sgr MPP and the two poorest SInt's -- to represent the background Sgr debris.
The uncertainty over which populations to use (and the relative strengths
of each) introduces some uncertainty into our distance measures.  This uncertainty is estimated to be around 0.05 magnitudes
based on a comparison of the distance moduli derived for various combinations of stellar populations.

The isochrones were first placed at M54's distance and NGC~6624's reddening, with the $A_{F606W}$ and $A_{F814W}$  extinction
coefficients set to those of Sirianni et al. (2005).  We then varied the distance and reddening
until the Sgr isochrones overlapped the
composite background population created from the shifted and co-added color-magnitude diagrams of all five clusters.
To check internal consistency, we
then fitted the isochrones to the background population of each cluster CMD on an individual basis.  We recover the
relative shifts between the different background features to a consistency of approximately 0.03 magnitudes.

\begin{center}
\begin{figure}[ht!]
\includegraphics[angle=0,scale=0.63]{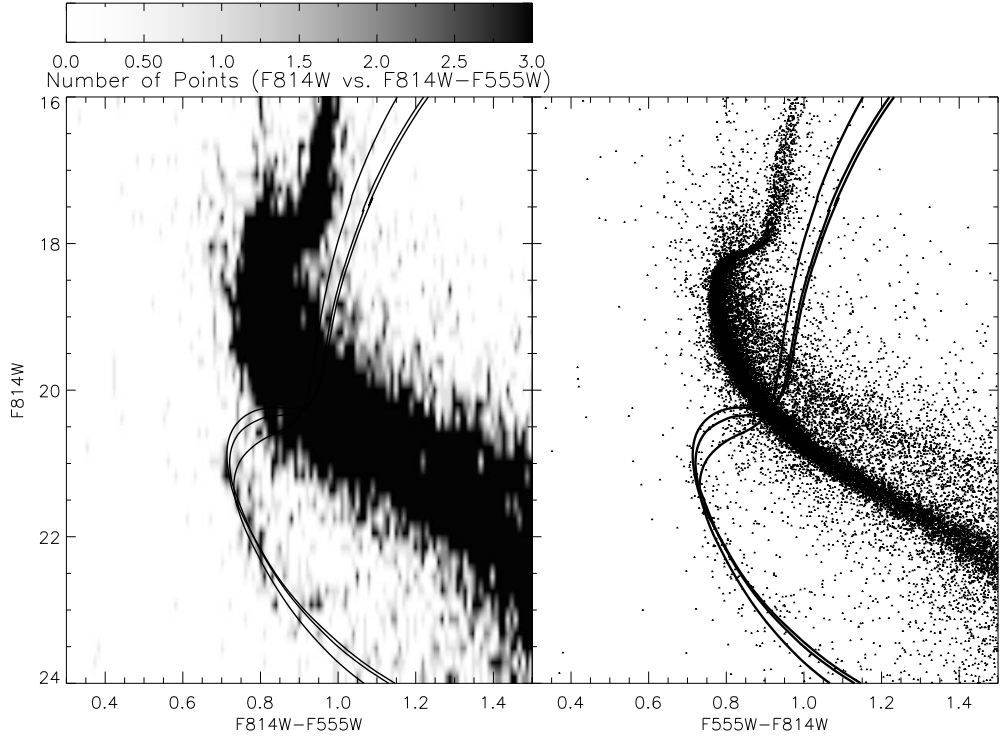}
\figcaption[f6.eps]{Constraining the background Sgr feature in the NGC~6624 bulge cluster field.  The left panel shows
a logarithmic Hess diagram with the scale stretched to provide maximum contrast for the Sgr background feature.  This has the effect
of saturating the foreground cluster.  The right
panel shows the unbinned CMD.  The isochrones are
an overlay of the adopted SInt and Sgr MPP populations used to fit the HST/ACS CMD of the Sgr core in Paper IV, shifted to the reddening and distance of
the Sgr feature from the field as given in Table 
\ref{t:bgtab}.\label{f:sgrbg1}}
\end{figure}
\end{center}

\begin{center}
\begin{figure}[ht!]
\includegraphics[angle=0,scale=0.63]{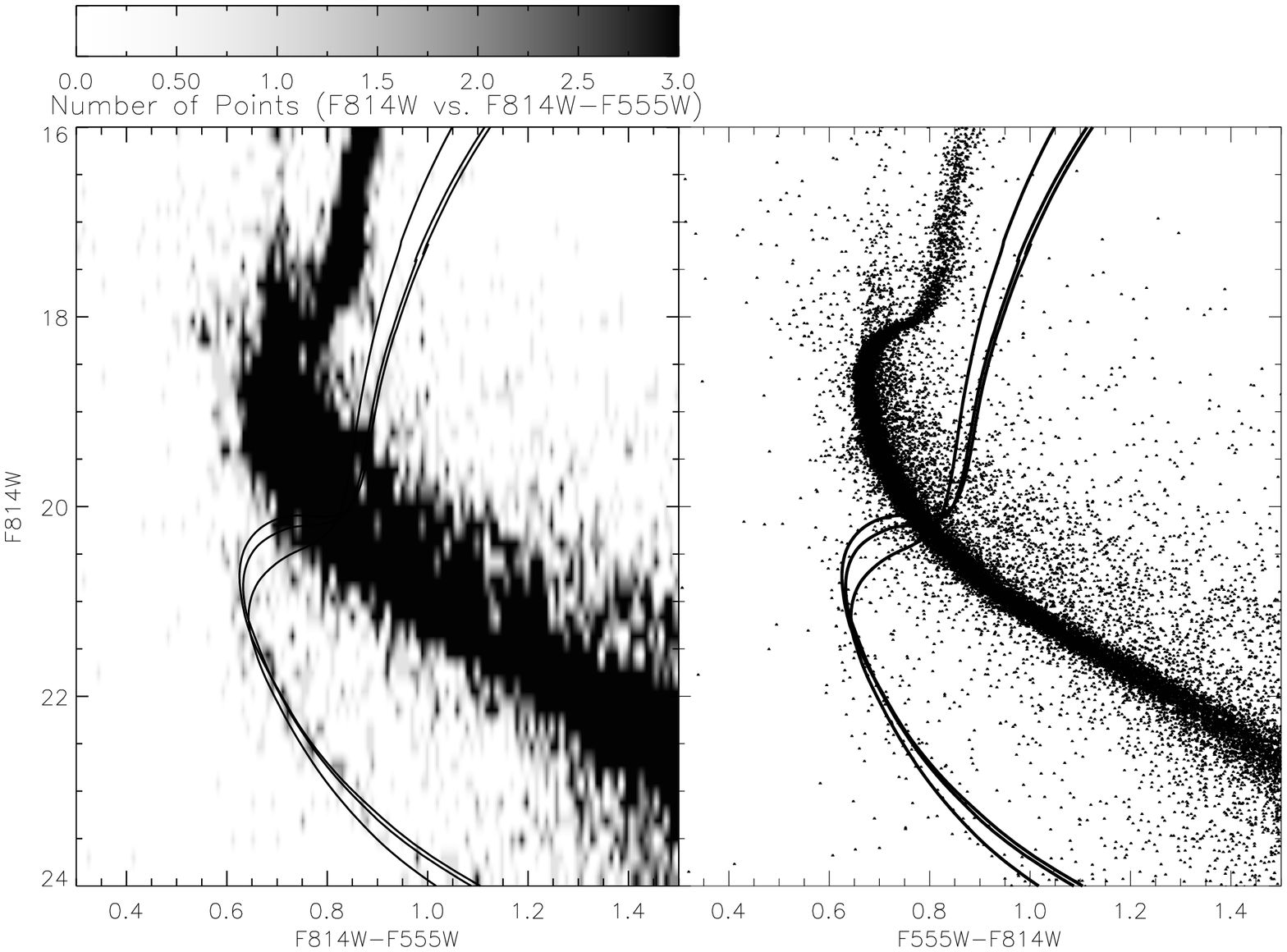}
\figcaption[f7.eps]{Constraining the background Sgr feature in the NGC~6637 bulge cluster field. See Figure \ref{f:sgrbg1} for description.
\label{f:sgrbg2}}
\end{figure}
\end{center}
\begin{center}
\begin{figure}[ht!]
\includegraphics[angle=0,scale=0.63]{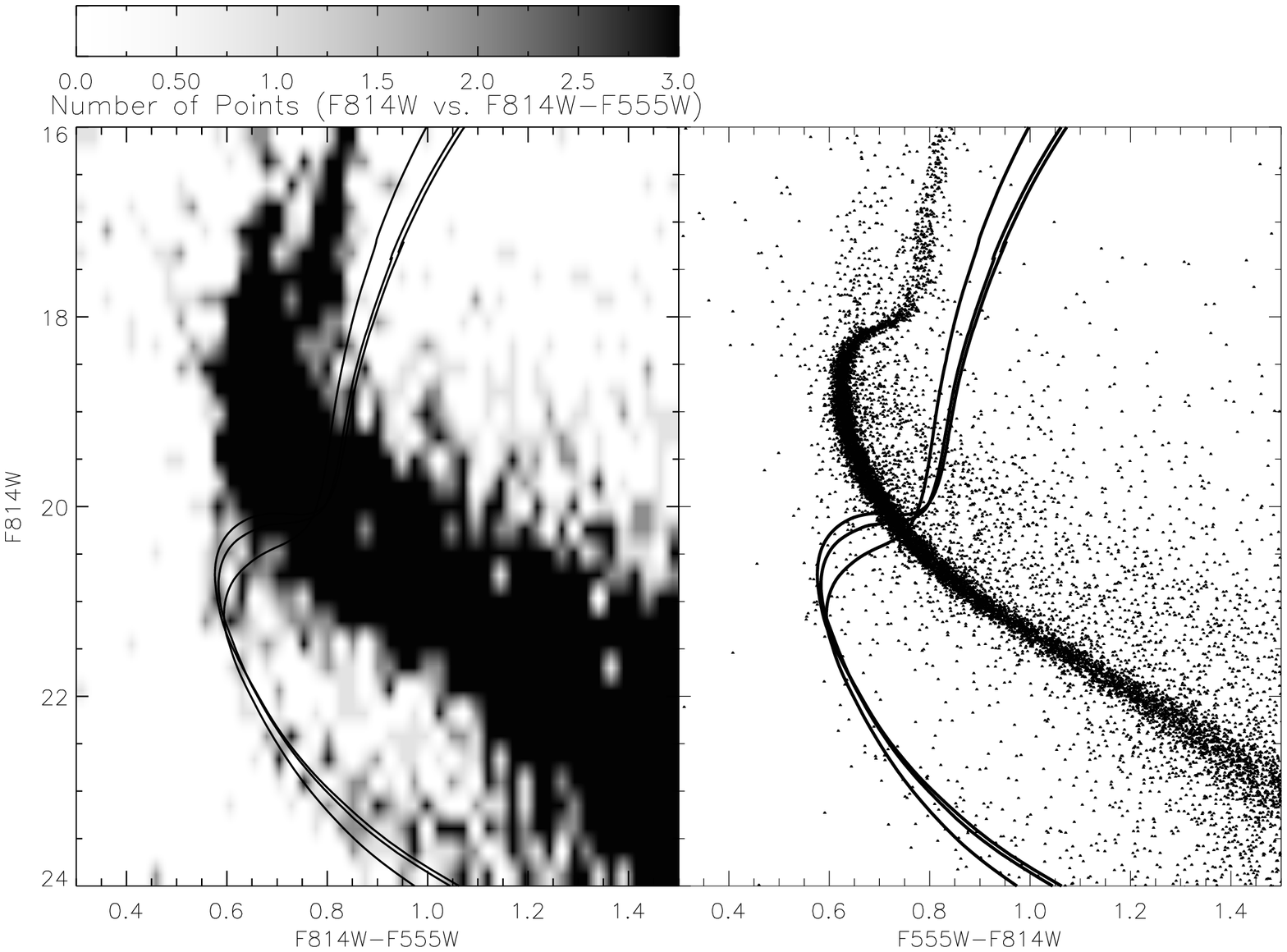}
\figcaption[f8.eps]{Constraining the background Sgr feature in the NGC~6652 bulge cluster field. See Figure \ref{f:sgrbg1} for description.
\label{f:sgrbg3}}
\end{figure}
\end{center}
\begin{center}
\begin{figure}[ht!]
\includegraphics[angle=0,scale=0.63]{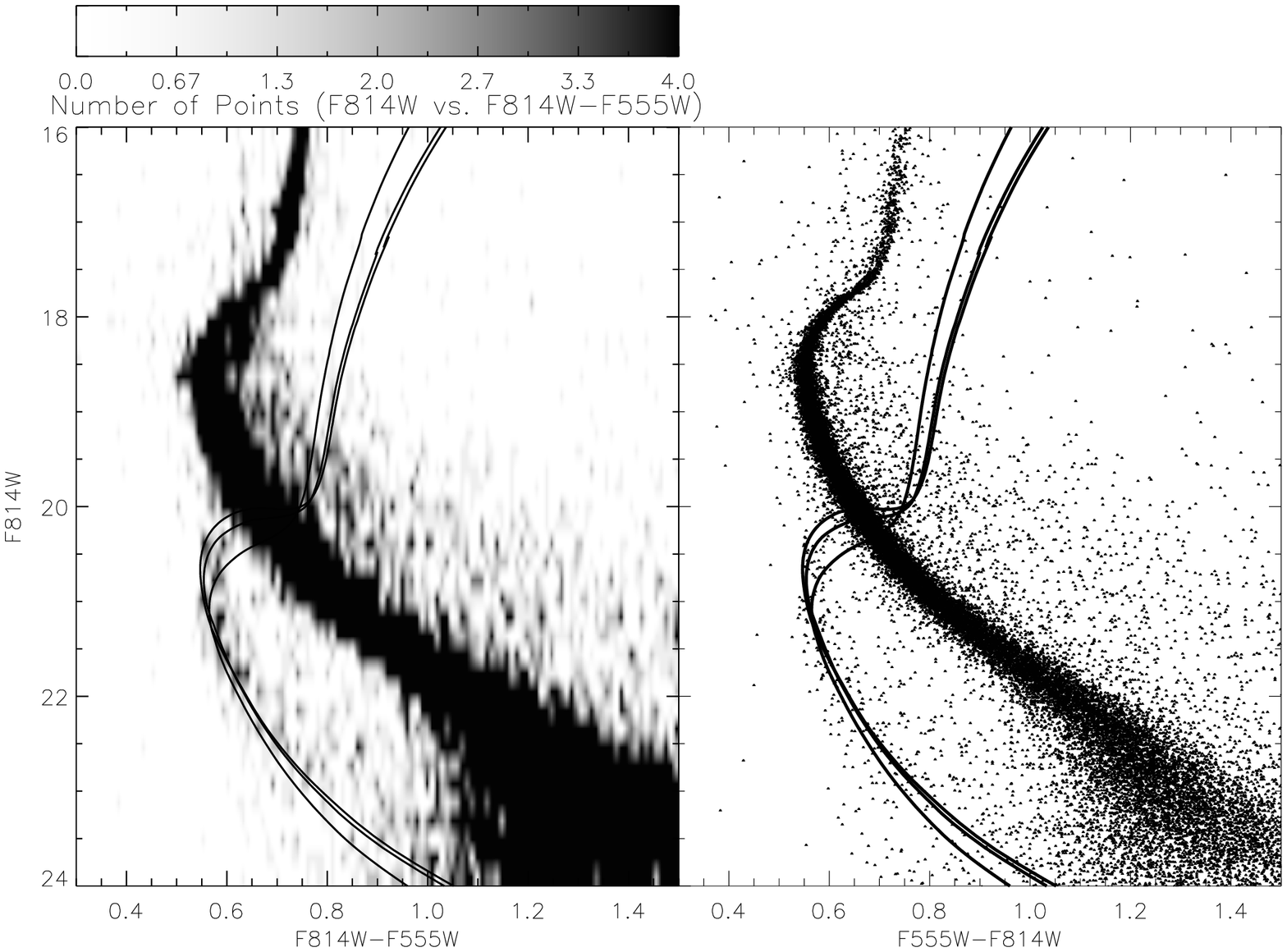}
\figcaption[f9.eps]{Constraining the background Sgr feature in the NGC~6681 bulge cluster field. See Figure \ref{f:sgrbg1} for description.
\label{f:sgrbg4}}
\end{figure}
\end{center}
\begin{center}
\begin{figure}[ht!]
\includegraphics[angle=0,scale=0.63]{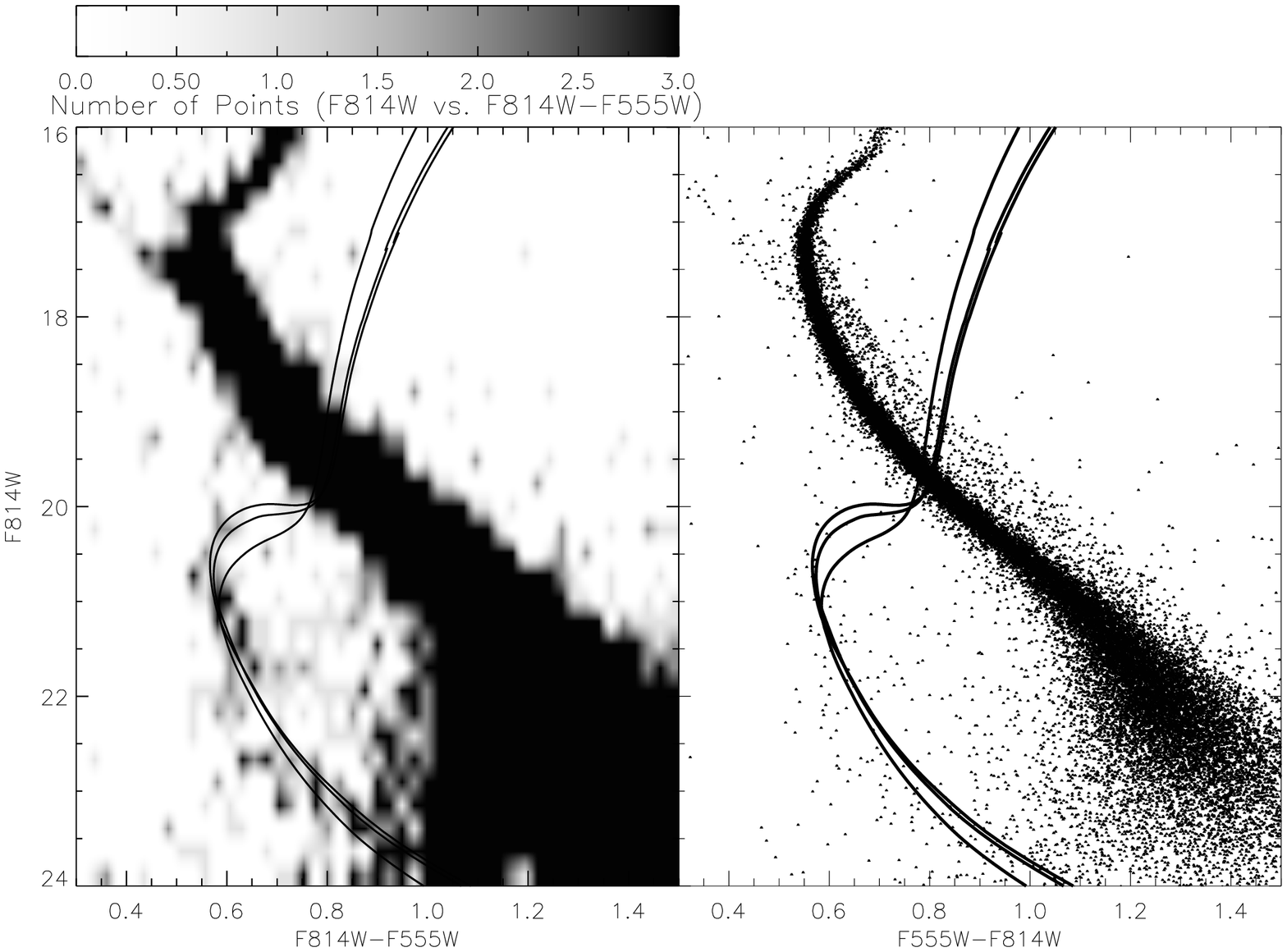}
\figcaption[f10.eps]{Constraining the background Sgr feature in the NGC~6809 bulge cluster field. See Figure \ref{f:sgrbg1} for description.
\label{f:sgrbg5}}
\end{figure}
\end{center}

Figures \ref{f:sgrbg1} through \ref{f:sgrbg5} show both the Hess diagrams of the CMDs and the unbinned CMDs of the foreground clusters with 
the fitted Sgr isochrones overlayed.  The parameters of the fits are listed
in Table \ref{t:bgtab} and include coordinates in the Sgr system, the distances and reddenings to the foreground
clusters, and the distances and reddenings to the Sgr CMD features.
$E(B-V)$ values are calculated from $E(F606W-F814W)$ using the coefficients of Paper I.
We estimate, for each field, that the {\it relative} distance modulus uncertainty is 0.05 magnitudes and the reddening uncertainty
is 0.01 mag.  The absolute distance uncertainty is larger, depending on the absolute calibration of the isochrones.
Table \ref{t:bgtab} also lists a value for 
$N_{\rm Sgr}$, 
a crude measure of the strength of the Sgr feature.  It is
simply the number of stars in the magnitude range $21 < F814W_0 < 23$ that are within 0.1 mag (quadrature of color and $F814W$ magnitude) of one of the three most metal-poor
Sgr isochrones.  The final column lists the linear distance to the Sgr CMD features in kpc.

\begin{center}
\begin{deluxetable}{lcccccccc}
\tabletypesize{\tiny}
\tablewidth{0 pt}
\tablecaption{Bulge Clusters With Sgr Background Features\label{t:bgtab}}
\tablehead{
\colhead{Cluster} &
\colhead{$\Lambda_{\odot}$ (deg)} &
\colhead{$B_{\odot}$ (deg)} &
\colhead{$(m-M)_{0,{\rm clus}}$} &
\colhead{$E(B-V)_{\rm clus}$} &
\colhead{$(m-M)_{0,{\rm Sgr}}$} &
\colhead{$E(B-V)_{\rm Sgr}$} &
\colhead{$N_{\rm Sgr}$} &
\colhead{D$_{\rm Sgr}$ (kpc)}}
\startdata
\hline
NGC~6624 & 353.4 & 2.8 &  14.58  &   0.26 &  17.33 & 0.31 & 155 & 29.2$\pm$0.7\\ 
NGC~6637 & 355.4 & 4.3 &  14.75  &   0.17 &  17.35 & 0.22 & 113 & 29.5$\pm$0.7\\ 
NGC~6652 & 356.5 & 4.8 &  14.84  &   0.12 &  17.42 & 0.17 &  62 & 30.5$\pm$0.7\\ 
NGC~6681 & 357.9 & 3.7 &  14.87  &   0.10 &  17.42 & 0.14 & 222 & 30.5$\pm$0.7\\
NGC~6809 &  9.7  & 0.8 &  13.67  &   0.12 &  17.34 & 0.16 &  84 & 29.4$\pm$0.7\\
\hline
\enddata
\end{deluxetable}
\end{center}

As may be seen, the distances of the Sgr CMD features are similar to those of the Sgr core and associated globular clusters.
All of the Sgr CMD background features show a slightly increased reddening compared to their respective foreground clusters of 0.02-0.05 magnitudes in $E(B-V)$.
On average, the Sgr CMD features are best fit with 0.04 magnitude additional reddening before that of the foreground cluster and 0.02 magnitudes more reddening
than the values estimated from the reddening maps of Schlegel et al. (1998).  The additional color shift compared to the foreground
cluster could reflect additional reddening
along the line of sight.  The foreground clusters are all within 5-9 kpc of the Sun and
all have Galactic $Y$ and $|Z|$ coordinate values of less than 1 and 2 kpc, respectively (with $XYZ$ centered on Sun, $Y$ defined along the direction of Galactic rotation
and $Z$ defined as distance above the plane). While the $|Z|$ distances are large compared to the scale height of the dust at the solar radius ($\sim$134 pc, 
Marshall et al. 2006), the dust scale height is expected to be larger in the Galactic bulge.
Even a small amount of additional dust from the bulge could produce the required additional reddening (although we note that all five clusters are close
to the total column reddenings given in Schlegel et al. 1998).

\begin{center}
\begin{figure}[ht!]
\includegraphics[angle=0,scale=0.63]{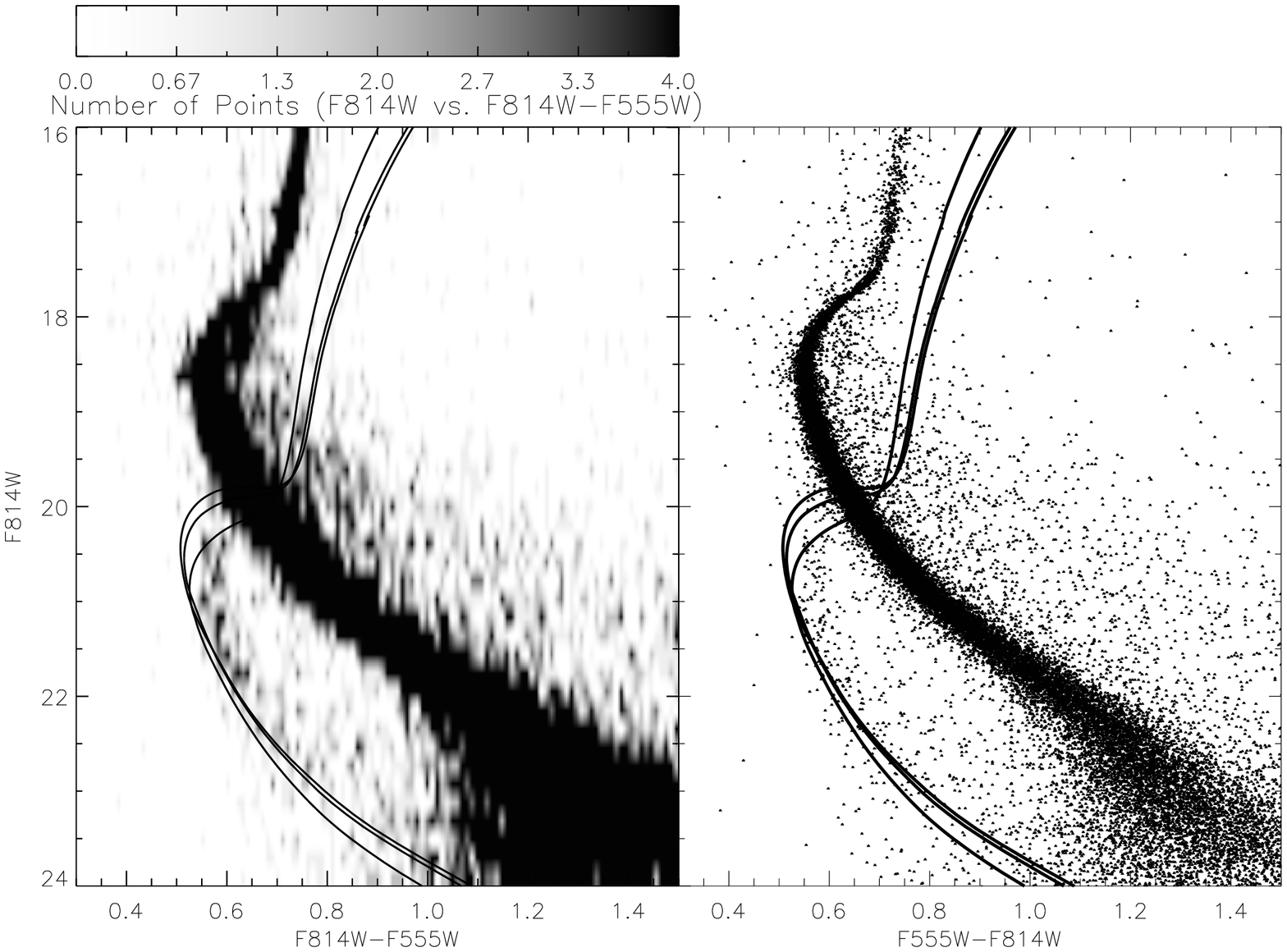}
\figcaption[f11.eps]{A modified version of Figure \ref{f:sgrbg4} in which the Sgr isochrones are set to the reddening of the foreground cluster and the
distance of M54.  Note that the isochrone miss the MSTO badly without the additional reddening.
\label{f:sgrbad}}
\end{figure}
\end{center}

The additional shift in color is required to provide an adequate 
isochrone fit to the background features.  Figure \ref{f:sgrbad} shows NGC~6681 with the Sgr isochrones overlayed at the
canonical Sgr distance and NGC~6681's foreground reddening. The isochrone MSTO is notably brighter and bluer than the observed MSTO and the MS of
the isochrone does not overlap the observed MS at all. Shifting the isochrone to the observed MS with a pure magnitude shift would require the Sgr feature
to be 0.3 magnitudes closer along the line of sight and do a poor job of recreating the slight upward turn of the brighter portion of the MS just before the turnoff.

An alternative explanation for the additional measured reddening is that we have misinterpreted as reddening a change
in the turnoff color over the extended distribution of Sgr.  Such a difference in turnoff
color would require the extended Sgr population be either older or more metal-rich than the Sgr core population.  The latter is unlikely, given
the known metallicity gradients.  However, the former -- an age difference --
remains a possibility.  Our analysis explicitly includes an age differential in Sgr by using only the older isochrones from Paper IV in our
analysis of the background features.  Paring
down the Sgr background population to only the oldest SInt population and Sgr MPP from Paper IV would move the background features closer
by approximately 0.05 magnitudes in distance modulus.  Stripping the Sgr stream population down to only Sgr MPP would move the background features closer
by another 0.14 magnitudes.  However, neither of these changes would remove the readily apparent reddening difference between the foreground bulge
cluster and background Sgr feature, because the older populations are also more metal poor and therefore have a bluer MS.  Moreover, a single population
would result in a Sgr background feature far narrower in color-magnitude space than that seen in the CMDs -- either in our study or the recent HST study of
Pryor et al. (2010) -- and conflicts with the abundance distribution observed by Frinchaboy et al. (in prep).
Finally, stripping down the Sgr field population to its oldest components would result in a metallicity for the extended Sgr dSph that
is {\it lower} than that measured in the extended streams by Alard et al. (2001) and Chou et al. (2007).  We therefore find it unlikely
that the difference in apparent foreground reddening is a mistaken analysis of a difference in stellar populations.

A final explanation is that systematic errors in the models and synthetic (or even empirical) colors is masking itself as a difference in reddening, i.e.,
that we are measuring a discrepancy in the isochrones rather than a discrepancy in the reddening.  We find
this unlikely, given the range of abundances in the foreground clusters (-0.44 to -1.94) and the success the isochrones have had in reproducing the
color-magnitude shapes of the full sample of clusters in Paper IX.

In the end, however, our analysis is based on the measures of Sgr features relative to each other, not to the foreground clusters.  So while a problem
in reddening or photometry may change the absolute values of our measurements, it will have no effect on the relative measures.  This is particularly
germane to NGC~6681, which has the strongest Sgr feature and a color shift {\it identical} to that of the M54 central field.  We are comparing
the same isochrones to the same populations with the same color locus, minimizing any differential problems with our calibration or analysis.

\section{The Three-Dimensional Orientation of the Sagittarius dSph}
\label{s:3dsgr}

Previous investigations of Sgr establish a baseline against which to compare and contrast our new Sgr distance and density measures.  In
this section, we compare the new CMD detections to both previous observational data and theoretical models.

\begin{center}
\begin{figure}[ht!]
\plotone{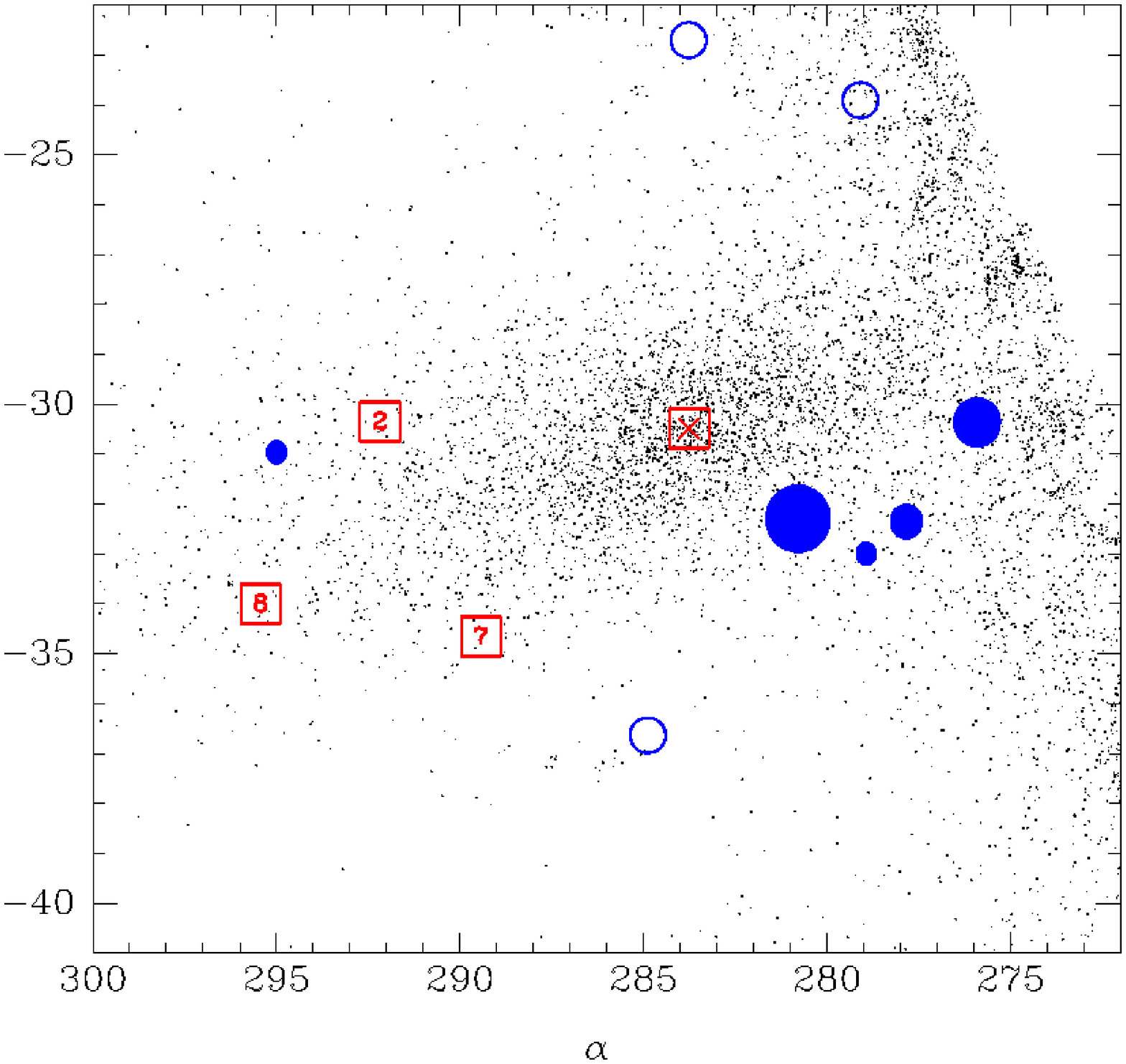}
\figcaption[f12.eps]{The ACS Survey detections of Sgr clusters and Sgr field stars seen in the ACS globular cluster survey fields as
background debris compared to the M-giant map in the area around the Sgr dSph by Majewski et al. (2003). Red squares
indicate Sgr globular clusters while filled blue circles are Sgr detections as background features to Milky Way globular clusters,
with point size indicating the strength of the
Sgr detection. Open blue circles are three clusters in the ACS survey program that do not have Sgr CMD features.
\label{f:2masscomp}}
\end{figure}
\end{center}

Figure \ref{f:2masscomp} shows the positions and densities of our Sgr detections compared to the plot of M-giants from M03.  The open squares
represent the Sgr clusters while the solid points are the background features, with the size reflecting the relative density of the Sgr detections
as measure by $N_{\rm Sgr}$ in Table 2.  Open circles represent ACS survey targets that do not have Sgr CMD features in them.  We have not used
the M-giant distances for our analysis because they are highly uncertain (typical $\sigma_d/d\sim0.2-0.25$, M03), are on a different distance scale than
the ACS photometry and are calibrated to a single color-magnitude relation that may not adequately reflect the abundance gradient within Sgr that must
be accounted for to precisely probe Sgr's 3-D shape.

\begin{center}
\begin{figure}[ht!]
\plotone{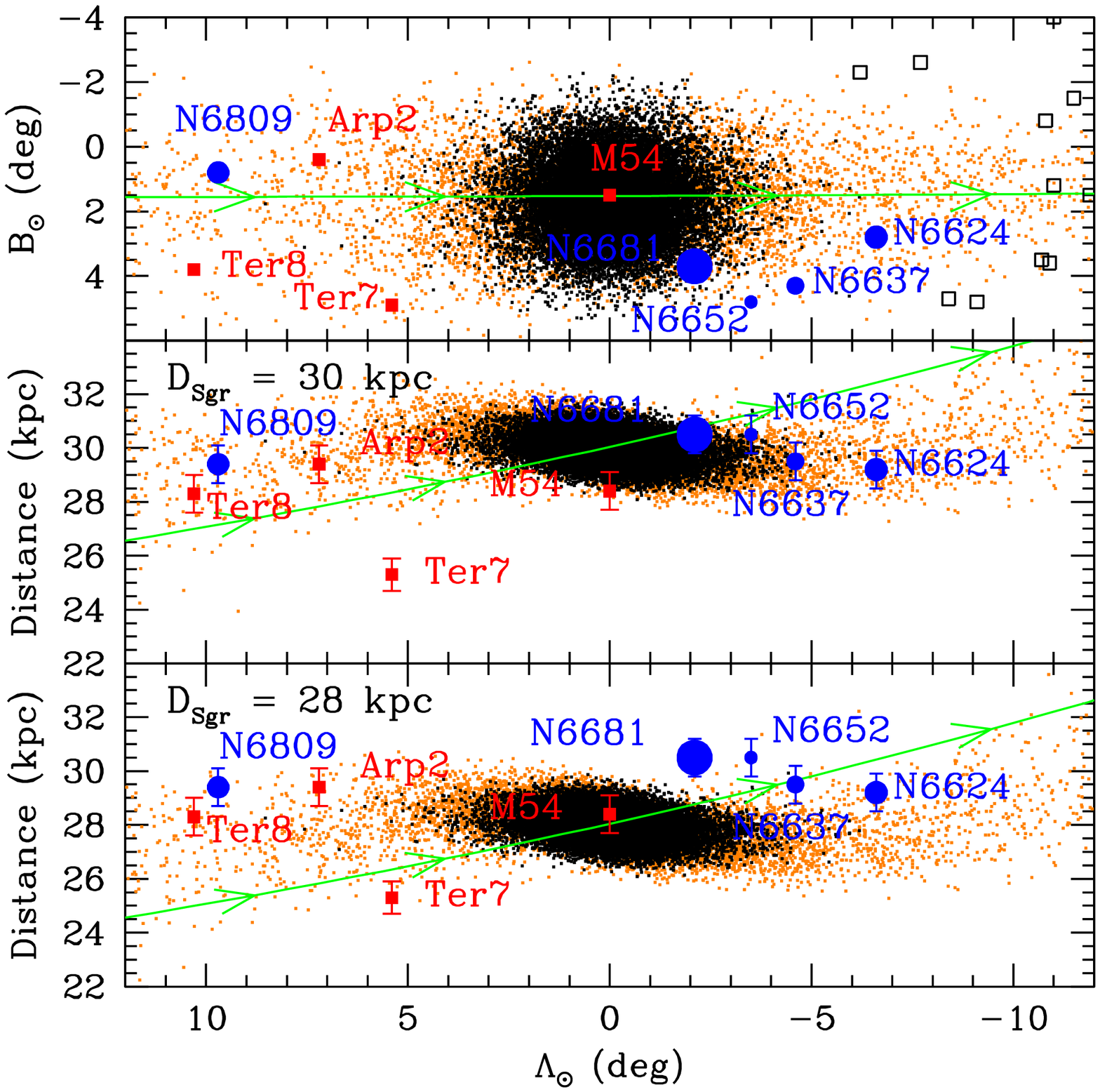}
\figcaption[f13.eps]{Distances and Sgr orbital latitude $B_{\odot}$ plotted against orbital longitude $\Lambda_{\odot}$ for the Sgr
clusters (red squares) and Sgr CMD background features (blue circles) as well as globular clusters not included in the ACS Survey of Galactic
Globular Clusters (open squares).
The size of the Sgr main sequence feature points is scaled to the number of Sgr member stars ($N_{\rm Sgr}$) listed in Table \ref{t:bgtab}.
These are overlaid against the LM10a model of the Sgr tidal streams: Black points represent particles 
current bound to the dSph, while orange points have recently become unbound from the dwarf.  The green line represents the orbital path of 
Sgr, with arrows representing its direction of motion. The middle panel shows the distance profile for a simulation in which the distance
to Sgr has been set to 30 kpc.  The bottom panel shows the distance profile for a simulation in which the distance to Sgr has been set to 28 kpc.
\label{f:sgrsim}}
\end{figure}
\end{center}

Figure \ref{f:sgrsim} shows the positions and distances of our Sgr clusters and background features
compared with two variations of the LM10a $N$-body model of Sgr disrupting in the Milky Way potential, one with Sgr at a distance of 28 kpc along the
line of sight and one with Sgr at 30 kpc along the line of sight.
We have made a minor adjustment to the models to center them at precisely
$\Lambda_{\odot} = 0^{\circ}$ and $r = 28$ or 30 kpc; although LM10a constrained their model to lie at $\Lambda_{\odot} = 0^{\circ}$ and $r = 28$, numerical limitations
of the $N$-body method caused their model Sgr dSph to overshoot the correct position at the current epoch by $\sim 2^{\circ}$ (corresponding to $\sim 3$ Myr)
along its orbit.
While this slight mismatch was unimportant for the analysis of LM10a (which focused on the properties of the Sgr tails at large angular
separations
from the dwarf), it is significant with respect to our present study of the Sgr core.  We therefore manually wound the simulations
back $\sim 3$ Myr to place the Sgr core at the fiducial location.  

\subsection{The Breadth of the Sgr Stream}
\label{s:sgrbreadth}

The sky distribution of our Sgr detections (Figure \ref{f:2masscomp} and top panel of Figure \ref{f:sgrsim}) shows that the clusters
and background features span the observed and modelled width of the stream.
While we will address the 
distance measures in \S\ref{s:dlambda}, the simple {\it presence} of Sgr debris is
a useful constraint on Sgr's disruption.  The detection of Sgr over such a large range in $B_{\odot}$ is consistent
with the breadth of the stream
depicted in M03 as well as the relatively
high Sgr core mass used in the LM10a model.  A smaller Sgr mass would result in narrower streams that would be
inconsistent with the spatial distribution shown in the figures.  For example, reducing the Sgr mass by a factor of two would
leave the NGC~6652 and NGC~6637 Sgr features well outside the tidal stream (while also giving a stream velocity dispersion inconsistent with
observations -- see Figure 4 of LM10a).

Expanding the range of Sgr pencil beams, especially across the face of the leading arm, would provide tighter constraints on the three-dimensional
orientation of Sgr and the disruption models.
A number of globular clusters are within the field depicted in Figure \ref{f:sgrsim} (open squares in upper panel)\footnote{The clusters Djorg2, NGC~6522, NGC~6528, NGC~6540, NGC~6544, NGC~6553, NGC~6558,
Terzan 12, NGC~6569, NGC~6626 and NGC~6638 would lie within the field of Figure \ref{f:sgrsim}.} and could potentially provide this additional
constraint.  Of the eleven clusters that could provide additional
information, three have not been observed with HST/WFPC2 or HST/ACS.  The remaining eight appear in Piotto et al. (2002), but the HST/WFPC2 data
are not deep
or precise enough to detect the faint Sgr stream.  NGC~6522, NGC~6528, NGC~6544, NGC~6553 and NGC~6558 have been observed with ACS
but the data are not as deep as that of the ACS Treasury Program and 
lack the photometric precision to delineate the faint Sgr sequence.  Published CMDs of NGC~6553
(Feltzing \& Gilmore 2000; Zoccali et al., 2001; Bealieu et al. 2001) and NGC~6528 (Brown et al. 2005) do not clearly show the stream.
Our deep precise HST imaging is the first to clearly discern these faint sequences.

Future deep observations of the above-named clusters -- or any field within the Sgr stream --
are recommended.  Although the clusters marked on Figure \ref{f:sgrsim} are further away from the Sgr core -- and would therefore have lower Sgr
debris densities than even NGC~6652 -- the mere presence or absence of the Sgr stream in these fields would help
constrain the models
of the disrupting core by providing hard limits on the breadth of the stream in $B_{\odot}$.
This is particularly true of NGC~6626/NGC~6638 and NGC~6558/NGC~6569, which occupy useful locations in the
Sgr coordinate system ($(\Lambda_{\odot},B_{\odot}$) $\sim$ (353,-2) and $\sim$ (351,5), respectively).

The nearest clusters in our ACS program to the Sgr core, other than those in Table 2, are NGC~6656, 6717 and 6723 (marked as open circles in Figure \ref{f:2masscomp}).
However, none of these clusters show any indication of Sgr debris.  This is expected given the relatively high
$B_{\odot}$ positions of the clusters (-4.1, -6.1, and +7.5, respectively).

\subsection{The Density Distribution}
\label{s:density}

\begin{center}
\begin{figure}[ht!]
\plotone{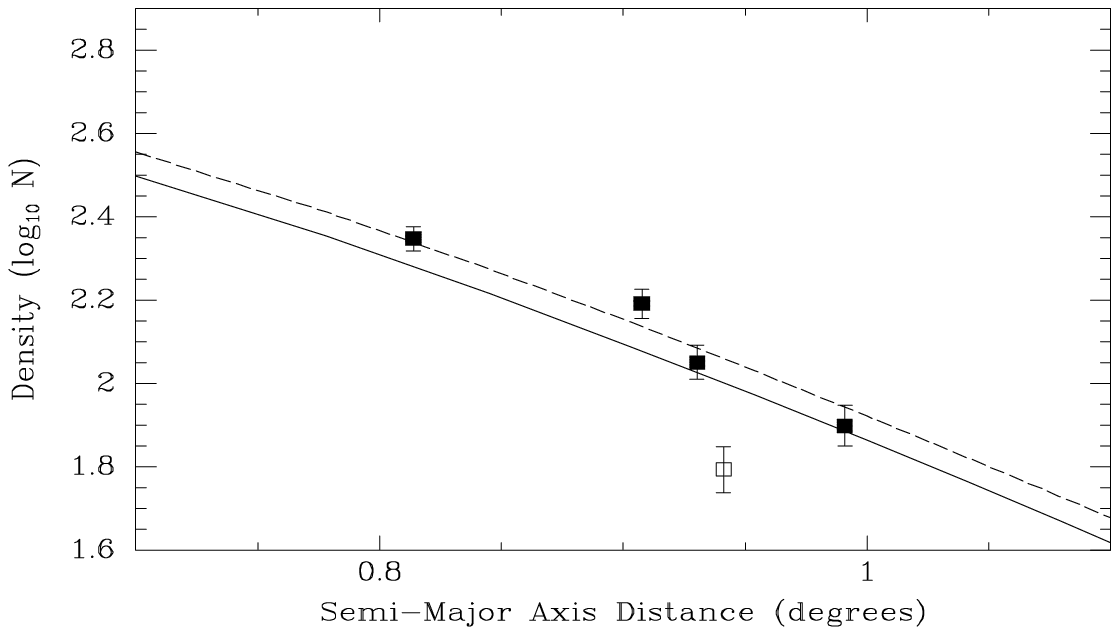}
\figcaption[f14.eps]{Comparison of the density of Sgr background debris to exponential profile
described in Table 1 of M03. The
profile has been arbitrarily scaled to matched the mean density of the Sgr main sequence stars. NGC~6652 (open square) has the lowest
density of the data points. The lines show the best fit density profile including NGC~6652 (solid line) and excluding it (dashed line).
\label{f:sgrdencomp}}
\end{figure}
\end{center}

Figure \ref{f:sgrdencomp} compares the density of Sgr main sequence stars ($N_{\rm Sgr}$) measured in our five ACS fields against the power
law+core density model fit to M giants from Table 1 of M03.  The
relative density level has been scaled to minimize the $\chi^2$ of the comparison without altering the core radius, ellipticity or
index of the M03 power law model.

Four of the fields show consistent relative densities -- within 1-3 $\sigma$ of the prediction.  NGC~6652, however, represented in
Figure \ref{f:sgrdencomp} by an open square,
is significantly (7 $\sigma$) off from the prediction, despite its proximity to NGC~6637, which is at a similar radial distance and
is the closest to the model prediction.  Removing
NGC~6652 from the fit would reduce the $\chi^2$ from 31 to 3.9 (dashed line in Figure \ref{f:sgrdencomp}).

We explored whether a change to the M03 model could bring the NGC~6652 data point back into line.
Five data points are insufficient to constrain either a King or exponential density model.  However, we compared the data
points to a series of ``toy models" using the power law and core model from M03 with a variety of core radii, power law indices and ellipticities.  The data
were insensitive to the core radius and favored a slightly shallower power law index than that of M03 (2.4-2.5, compared to M03's 2.6).  We also found that the data favored
a slightly higher ellipticity than M03's 0.62 of 0.67 (with NGC~6652) or 0.65 (without).  However, even these changes only marginally improved the fit of the model,
with and without NGC~6652, to $\chi^2$ values of 29 and 3.6, respectively.  No simple exponential model could bring NGC~6652 into line with the other background features.

What could be the cause of the discrepancy in NGC~6652?  Completeness is not an issue; the artificial star tests described in Paper V show that
all five clusters have 90-95\% completeness levels to $F814W=23$.  It is unlikely that any of our fields suffer from significant
contamination from the foreground Galactic  stars, despite their proximity to the midplane.  CMDs for the non-Sgr detection fields of NGC~6656, 6717 and 6723,
at comparable Galactic $b$'s, show
less than ten stars in the CMD box from which $N_{\rm Sgr}$ is calculated, which suggests a contamination level for NGC~6652 of less than 1/6 from
all sources and the Besancon model (Robin et al. 2003) predicts only 1-3 Galactic halo stars in the CMD region from which we calculate $N_{\rm Sgr}$.
We could be
suffering from photometric contamination from the foreground
clusters in some of our fields.  Some of the Sgr features -- that of NGC~6681's CMD in particular -- are 
close to the MS of the foreground cluster and it is possible that they suffer from some photometric contamination that
inflates their densities slightly.  However, it would take a significant (40\%) amount of hypothetical foreground contamination
in the measured Sgr density in the other four clusters to make them consistent with NGC~6652.

If the density dropoff in NGC~6652 is real, it would indicate that while the 
density profiles fit in M03 provide a sound broad description of Sgr's density profile, the
narrow field of ACS is more susceptible to small variations in density, such as might be produced by isophotal twisting or a transition from the bound core of Sgr to the unbound
tidal stream (the contours of Ibata et al. (1995) hint at some irregularity in Sgr's inner structure).  Far more extensive data will be needed to determine if the measured NGC~6652 field density of Sgr MS stars
is simply a statistical anomaly or photometric error (albeit at the 7 $\sigma$ level) or an indication of a much more complex inner structure in Sgr.

Comparing the Sgr stellar densities to the $N$-body model shows a rough agreement.  NGC~6637 and 6652
are at large $B_{\odot}$ values while NGC~6809 and 6624 lie closer to the Sgr orbital plane.  NGC~6681, which has the highest
density, would be near the bound core of Sgr.  The model does not explain the low density we measure for NGC~6652.  
However, models with a greater variety of Sgr core shapes (e.g., {\L}okas et al. 2010) may be better able to reproduce the observed sky densities.

\subsection{The Distance-$\Lambda_{\odot}$-$B_{\odot}$ Relations}
\label{s:dlambda}

The $N$-body model of LM10a makes specific predictions about the relationship between sky position and line of sight distance for the disrupting
Sgr dwarf.  In the model, emerging debris in the trailing arm (positive $\Lambda_{\odot}$ values) is
slightly more distant that emerging debris in the leading arm (negative $\Lambda_{\odot}$).
This reflects the energy of the debris lost to the tails.  The leading
stream falls inside the orbital path and the angular momentum moves it ahead of
Sgr along the orbit.  The trailing
stream falls outside the orbital path and the angular momentum moves it behind Sgr.  Debris within five degrees of the Sgr core
should follow this trend.  According to the model, we should see an increase in distance with increasing $\Lambda_{\odot}$, albeit to a degree ($\sim$ 2 kpc)
that would be close to the {\it relative} distance uncertainties in our study (0.6-0.7 kpc assuming a relative distance modulus uncertainty
of 0.05 magnitudes).

At radii beyond five degrees, the distance-$\Lambda_{\odot}$ trend reverses because of the shape of Sgr's orbit (the green line in Figure \ref{f:sgrsim}).  The dSph
is moving away
from the Sun, so leading debris will be further away while trailing debris will be closer.  According to the model, this
orbital shape will result in a difference in line of sight distances of approximately 4 kpc over the $\pm$12 degree span of our survey.
However, the combination of the two dynamical effects -- mass loss mechanics
and orbital path -- should combine to make the distance to Sgr roughly constant over the $\Lambda_{\odot}$ interval covered
by our data, with perhaps a slight overall trend of increasing distance with decreasing $\Lambda_{\odot}$.

Our points are close to the prediction of the model and the sense of the distance measures -- either flat or slightly increasing distance with
decreasing $\Lambda_{\odot}$ -- is similar to that of the unbound debris.  However, with the exception of Terzan 7, the bottom panel of Figure \ref{f:sgrsim} shows
that our distance measures are consistently farther than predicted for a disrupting Sgr core at 28 kpc.  Increasing the Sgr distance
to 30 kpc, however, (middle panel of Figure \ref{f:sgrsim})
brings the distance measures into remarkably improved agreement with the model.  We explore the possibility of a larger Sgr distance further in \S4.4 and 4.5.

\subsection{M54 - Sgr Core or Chance Alignment?}
\label{s:m54stat}

Bellazzini et al. (2008, B08) have argued, from measurements of velocity dispersion profiles, that M54 is not
the core of Sgr but may have formed independently and plunged to the core of Sagittarius due to dynamical friction. Their dynamical
simulations indicate that this process would take only a few Gyr.  However, if M54 is liberated from its assumed status 
as the Sagittarius core, it need not necessarily be at the same distance along the line of site, as is assumed by B08.

If M54 were not precisely in the center of Sgr but were situated a kpc or two in front of or in back of the nucleus, 
then the Sgr distance
in Paper IV would be accurate for M54 itself because that distance was measured from the dominant M54 metal-poor main sequence.  However,
the stellar populations within the parent dSph might be at a different distance.  In Paper IV, we tied Sgr's distance to that of M54 because M54
dominates the main sequence regions of the CMD and we could not discern a distance discrepancy between the two.  The possibility of a distance difference
between M54 and Sgr was not addressed.

Our new distance measures allow an independent constraint on the distance to Sgr in the absence of stars from the overlapping globular cluster.
As noted above, the most striking aspect of Figure \ref{f:sgrsim} is that two of the three Sgr canonical clusters and all five of the Sgr main sequence
features lie at the most distant edge of LM10a's theoretical Sgr stream for a distance of 28 kpc.  Increasing the input model Sgr distance to 30 kpc produces
better agreement between the measured distances and the $N$-body model.
Arp 2, Terzan 8 and all five background features would be comfortably within the Sgr stream at the longer distance
(although Terzan 7 would be more discrepant).
In this scenario, M54 would lie at the near edge of the dSph core, possibly still within the bound portions of the satellite but displaced
from the center.

Could this apparent distance discrepancy be produced by something other than a genuine 
distance discrepancy between the Sgr core and M54?  
We consider five potential phenomena that could mimic a genuine discrepancy
in distance between Sgr and M54:

(1) Observational error could produce an offset in the photometry that disguises itself as an apparent offset in distance.  However, such a systematic error is unlikely.
The chief advantage of the ACS Survey is that the data are uniform, taken with the same instrument and filters over a short span of
time and reduced and processed through an identical pipeline.  For the distance difference to be a result of observational error, there would
have to be a previously undetected change in the HST-ACS camera that affects all of the cluster fields {\it except} those of M54 and Terzan 7 (or, alternately, 
{\it just} those of M54 and Terzan 7).

(2) We could be systematically over-estimating the distance of the Sgr main sequence features because of the small number of Sgr stars. Perhaps
some of the brighter Sgr populations are too sparse to be clearly detected in the background features, causing us to shift the group of three isochrones, including
the brighter ones, too faint in an effort to fit the data.
However, NGC~6681, which has the strongest Sgr background signal, yields one of the most distant measures at $(m-M)_0$=17.42, a $3\sigma$
discrepancy from the Paper IV Sgr distance (based on our estimate of 0.05 mag uncertainty in the relative distance).

(3) Related to (2) is the possibility that we have over-estimated the reddening, as discussed in \S \ref{s:background}.  However, as noted in that
section, all of our comparisons are relative.  Moreover, NGC~6681, which has the strongest Sgr background signal, has an identical color shift to the central
Sgr/M54 field but a large (3 $\sigma$) magnitude shift.  These discrepancies can not be accounted for by errors in the assumed foreground reddening.

(4) As noted above, our distance measures are sensitive to the assumed Sgr stellar populations used to constrain the distance to the background
features.  However, restoring the younger populations to our isochrones would result in a {\it longer} distance for the CMD background features.  Removing the intermediate age populations
results in shorter distances for the Sgr background features.  However, this results in a background Sgr population that is inconsistent with 
photometric measures of more dense areas of Sgr (see, e.g., Pryor et al. 2010) as well as spectroscopic surveys (Frinchaboy et al., in prep) that definitively
show an intermediate population consistent with our choice of isochrones. Moreover, the Sgr feature in NGC~6681 is very close to
the Sgr core, perhaps within the bound remnant.  It is unlikely that the background Sgr material in this field is much more metal-poor and/or older
than the Sgr debris in the M54 field, given the abundances in the extended Sgr core and stream measured by Alard et al. (2001), Chou et al. (2007, 2010),
Giuffrida et al. (2010) and Frinchaboy et al (in prep).

(5) Finally, the notion that we can measure the distance of the Sgr core from measurements taken along the extended profile of Sgr assumes
that we can use the LM10a model to compare the Sgr core distance to those measure along its tidal arms.
It is remotely possible that a Sgr core at the same distance as M54 could have extended tidal arms that are at the distances measured in the
other eight fields due to a different shape.  However, this too is unlikely.  It is difficult
to configure Sgr's geometry so that {\it both} tidal arms are further away than the Sgr core.

In the end, we conclude that the most likely explanation for the distance discrepancy between the Sgr debris and M54
is that M54 is, in fact, in front of the main body of Sgr.  The mean distance modulus of the Sgr clusters is 17.22, shorter than the
Paper IV distance of 17.27, mostly because of the short distance we measure for Terzan 7.  Dropping Terzan 7 sets the mean distance
modulus to 17.30. The CMD background features have a mean distance of 17.37 and a dispersion of 0.04, indicating a Sgr
distance modulus between 17.32 and 17.42 (assuming a 0.05 magnitude relative uncertainty), corresponding to a linear distance of 29-30 kpc.

Increasing the Sgr distance, however, might change the assumed Sgr populations.  The Sgr populations that we use to measure the distance of the
background main sequence features were constrained in Paper IV based on the assumption that they were at the same distance as M54.
If the Sgr populations were more distant, however,
this would alter the ages and metallicities of the Sgr populations described in Paper IV, which were based on the bluer MSTOs and RGBs of Sgr, rather
than the M54-dominated MS.  This could, in turn, change the distances measured for the background features from isochrone fitting.

\begin{center}
\begin{figure}[ht!]
\includegraphics[scale=0.63]{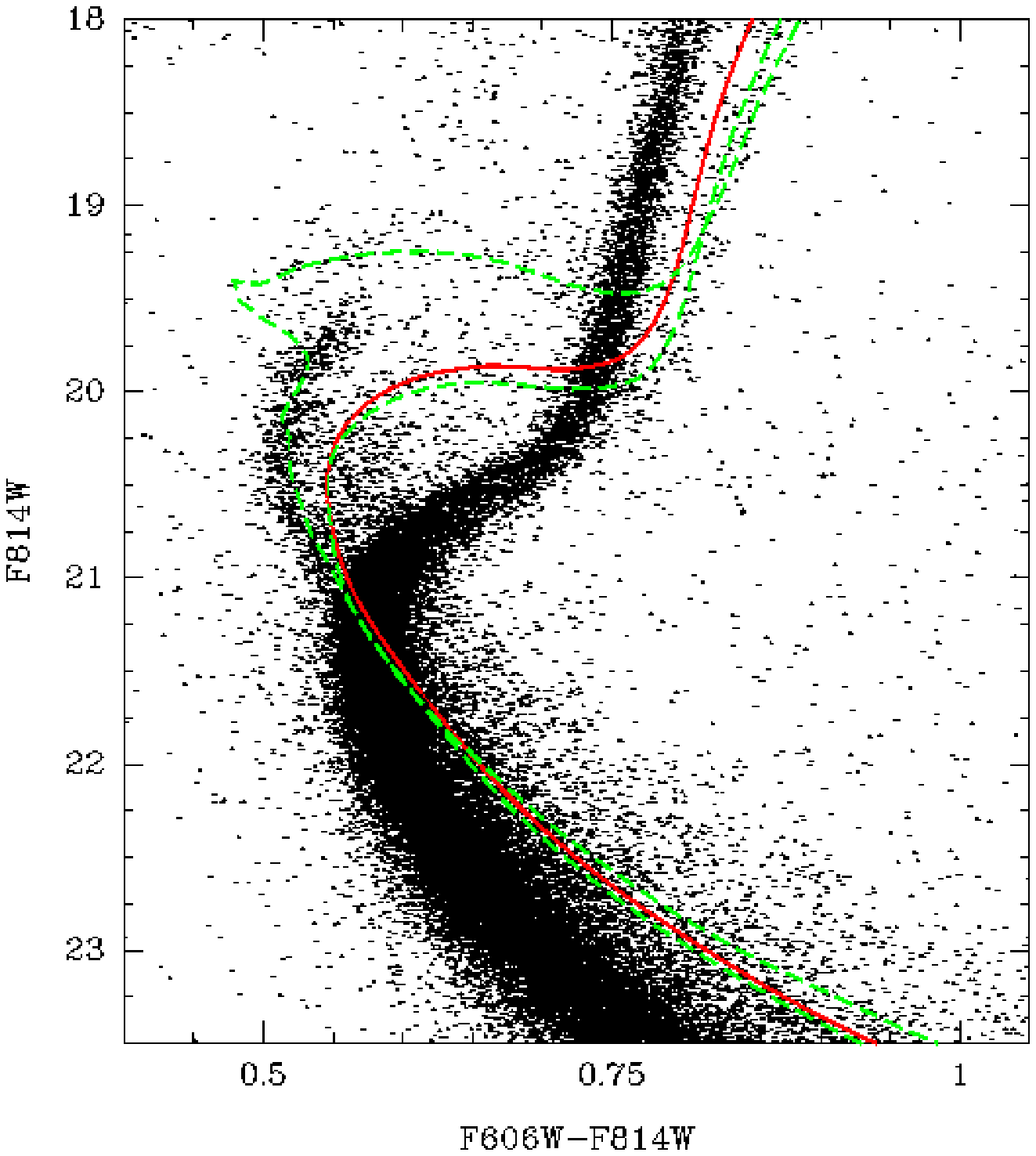}
\figcaption[f15.eps]{The effect of an increased Sgr distance modulus on the inferred stellar populations of M54/Sgr.  The solid red line
shows one of the SInt populations from Paper IV ([Fe/H]$= -0.5$, 6 Gyr). The lower dashed green line shows a similar population
([Fe/H]$= -0.4$, 5.5 Gyr) at the maximum distance modulus ($(m-M)_0=17.39$) implied by the bulge cluster Sgr CMD background features.  The
upper green dashed line shows the SYng population from Paper IV.  The revision we have made to the to isochrone color transformations for the younger
isochrones (see text in \S \ref{s:m54stat}) 
cancels out the increase in distance modulus, resulting in a nearly identical fit ([Fe/H]$= -0.2$, 2.8 
Gyr).\label{f:sgrmove}}
\end{figure}
\end{center}

However, our analysis indicates that
these changes to the measured Sgr stellar populations would be minor.  If we give Sgr a line-of-distance distance modulus of $(m-M)_0$=17.39, we
could easily align the intermediate isochrones
with the observed Sgr CMD features by decreasing their ages by 0.5-1 Gyr and perhaps increasing their abundances by to 0.1 dex (see Figure \ref{f:sgrmove}).  This would
have little effect on the CMD background feature distances.  Although the stellar population analysis of Paper IV was based on the MSTO and RGB, the
distances in this paper were measured based solely on main sequence fitting.  Altering the metallicity of the intermediate age
populations by 0.1 dex would make the RGB/MSTO consistent with the greater distance modulus but not significantly change
the main sequence luminosity more than our uncertainties.  The resultant distance measures would therefore be similar to those
derived above.  We cannot determine what effects the distance modulus change would have on the Sgr MPP since the characteristics population were assumed in Paper IV, not fit,
because of the dominance of the M54 MPP in the CMD.

As for the younger populations characterized in Paper IV, our assessment of them would be little affected by an increase in distance. The youngest population (SVYng) is
diffuse and its age/metallicity more uncertain than the distance scale.
The SYng population is better defined in the CMD. However, the increase in distance
would almost exactly be cancelled by a change in isochrone calibration.  Paper IV used semi-empirical colors for the
isochrones while this paper uses synthetic colors.
While the different color transformations produce similar isochrones for the older populations, there is some disagreement as
to the color and magnitudes of the younger populations. Given that the young populations are so far unconstrained
by spectroscopic metallicities, their ages and metallicities are not well-defined.
Nevertheless, the SYng population described in Paper IV still accurately reproduces the overall shape and color
of the SYng MSTO, as shown in Figure \ref{f:sgrmove}.  Of course, neither SVYng nor SYng are seen in the Sgr main sequence features, so they would not impact
the distance measures in the bulge cluster fields.

Based on our present information, it is possible that M54 is not at the center of Sgr.  Confirming or refuting this will require better constraints 
on Sgr's stellar populations
outside of the central M54 field and/or more distance measures along the face of the dSph.

\subsection{The Status of the Classical Sgr Clusters}
\label{s:otherclus}

One final problem related to using a longer distance for Sgr is the status of
the globular cluster Terzan 7.  Its distance of 25.4 kpc would be consistent with the canonical Sgr distance used by previous investigations. However,
it would be significantly (6 $\sigma$, assuming $\sigma_{m-M}=0.05$) different from the distance we measure for the other Sgr globular clusters (28.5 kpc)
and even more discrepant (7-9 $\sigma$) from the longer distance we argue to hold for the Sgr main body.

At first blush, it seems unlikely that Terzan 7 is only aligned with the Sgr dSph by chance.  Terzan 7's age and metallicity are
typical of the ``young halo" objects
traditionally associated with dSph galaxies.  Terzan 7 and Pal 12 are the only globular clusters known
to lack the Na-O anticorrelation that is ubiquitous in Galactic globular clusters (Carreta et al. 2010a).
Using previous distance measures, LM10b concluded that
Terzan 7 was unlikely to be aligned by chance based on the cluster's 
location, radial velocity and distance.  However, we decided to re-examine the latter point in light of our
new distance estimates.

To determine the status of Terzan 7, we re-ran the LM10a model and checked the status of all the Sgr member clusters for varying
Sgr distances.  If Sgr is at a distance of 28 kpc, 
Arp 2, Terzan 7 and Terzan 8 all
can be matched with individual test particles in the Sgr simulation.  In the simulation, the clusters have come unbound on the last pericenter (0.1 Gyr ago) and will move much
further (70-110 degrees) downstream before the next pericenter.  M54, of course, is in the center of Sgr in this simulation

Increasing the Sgr distance to 30 kpc and placing M54 2 kpc in the foreground changes this picture somewhat.  All four clusters match
test particles within the Sgr debris, but Terzan 7 requires a much more generous search tolerance.  Using the P3 statistic\footnote{P3 is the probability that an
artificial globular cluster inserted at random in the Galactic halo would by chance appear to match the Sgr stream as well as, or better than, the cluster in question.} from LM10b, we find that
Terzan 7's P3 value increases from 0.043 to 0.12 (compared to .001 for Terzan 8 and Arp 2).  
Given that $\sim 6$ such  false-positives expected at the level P3 $\sim 12$\%, it seems 
possible that Terzan 7 could be a chance alignment.  Given this now more troubling statistic, measuring an absolute proper motion for Terzan 7 is
critical to affirming its status as a member of Sgr.
If Terzan 7 is indeed a member, the distance discrepancy hints that Terzan 7 has a different
dynamical history than the other classical Sgr member clusters.

As for M54, the simulation with the cluster in front of the dSph indicates that M54 should sink to the center of Sgr via dynamical friction in
approximately 3 Gyr if the dSph has a cuspy core
consistent with the analysis of B08.  This would seem to argue against the idea that M54 is 1-2 kpc in front of the dSph.
However, if we assume a shallower density profile for Sgr, M54 may stall its descent near the core radius of $\sim 2$ kpc (M03).
Given the similarity between 
this distance and our measured M54-Sgr distance (2 kpc) this may already
have happened, although it would seem a remarkable coincidence that M54 has stalled precisely along our line of sight to Sgr.\footnote{In this scenario,
M54's radial velocity, at this point in its orbit, would still be very close to that of the core, consistent with the measurement of B08.}

\section{Discussion and Conclusions}
\label{s:conc}

Our study of the Sagittarius system using ACS-WFC photometry has identified parts of the Sgr core and debris trail
in the background of
five unrelated bulge globular clusters.  Assuming this background population to be consistent
with the intermediate-age and old populations identified in Paper IV, we derive distances to these features.

The combination of five lines of sight  to the Sgr main body as well as four Sgr member clusters allows nine precise and independent distance measures
to the center of the Sgr system.  The combined distances are roughly consistent with the Sgr distance measured in Paper IV for M54
($(m-M)_0$=17.27; $d$=28.4 kpc), which is slightly longer than the estimates based on the RGB tip (17.10; Monaco et al. 2004) and RR Lyrae
stars (17.19; Layden \& Sarajedini 2000; Kunder \& Chaboyer 2009).  However, the CMD background sequences are, in fact,
slightly more distant than the Paper IV estimate.  This may indicate that
M54 -- which is usually used as a proxy for measuring the distance to Sgr -- lies in the foreground of the dwarf and that
the true distance to Sgr is $\sim 2$ kpc larger than that derived in Paper IV. An increased
Sgr distance would be more consistent with the distances we measure for both the background features and two of the other Sgr
core globular clusters (Arp 2 and Terzan 8).  If M54 were 2 kpc closer along the line of sight, its distance would still be consistent
with the ``Sgr" distance measures made from RR Lyrae and RGB tip stars, measures that would be dominated by the M54 cluster.  Dynamical
models, both from B08 and LM10a, indicate that M54 would sink to the center of Sgr from dynamical friction in 2-3 Gyr. If it has not done so,
this may indicate a shallower density profile for Sgr or that M54 has a different dynamical history than assumed.

Terzan 7 is several kpc closer to us than the main body of Sgr and the other core Sgr clusters.  This raises the
possibility that Terzan 7 may not be part of Sgr or has a different dynamical history than that of the other globular clusters.
Precise absolute proper motions for the cluster would determine which of these scenarios is more likely by tracing the orbit of Terzan 7.

We find that the stellar densities for the Sgr features are consistent with the elongated Sgr shape described in M03, which the exception of an unusually
low density in the Sgr background feature of NGC~6652, which we cannot explain.  The relative distance measures across the face of Sgr are
also broadly consistent with expectations from current models of the Sgr
tidal disruption (see, e.g., LM10a) as long as the longer distance scale (30 kpc) for the Sgr system is adopted.
Our initial study indicates a substantial width to the emerging Sgr tidal tail, which
confirms the higher Sgr mass used in the LM10a model.

\begin{center}
\begin{figure}[ht!]
\includegraphics[scale=0.7]{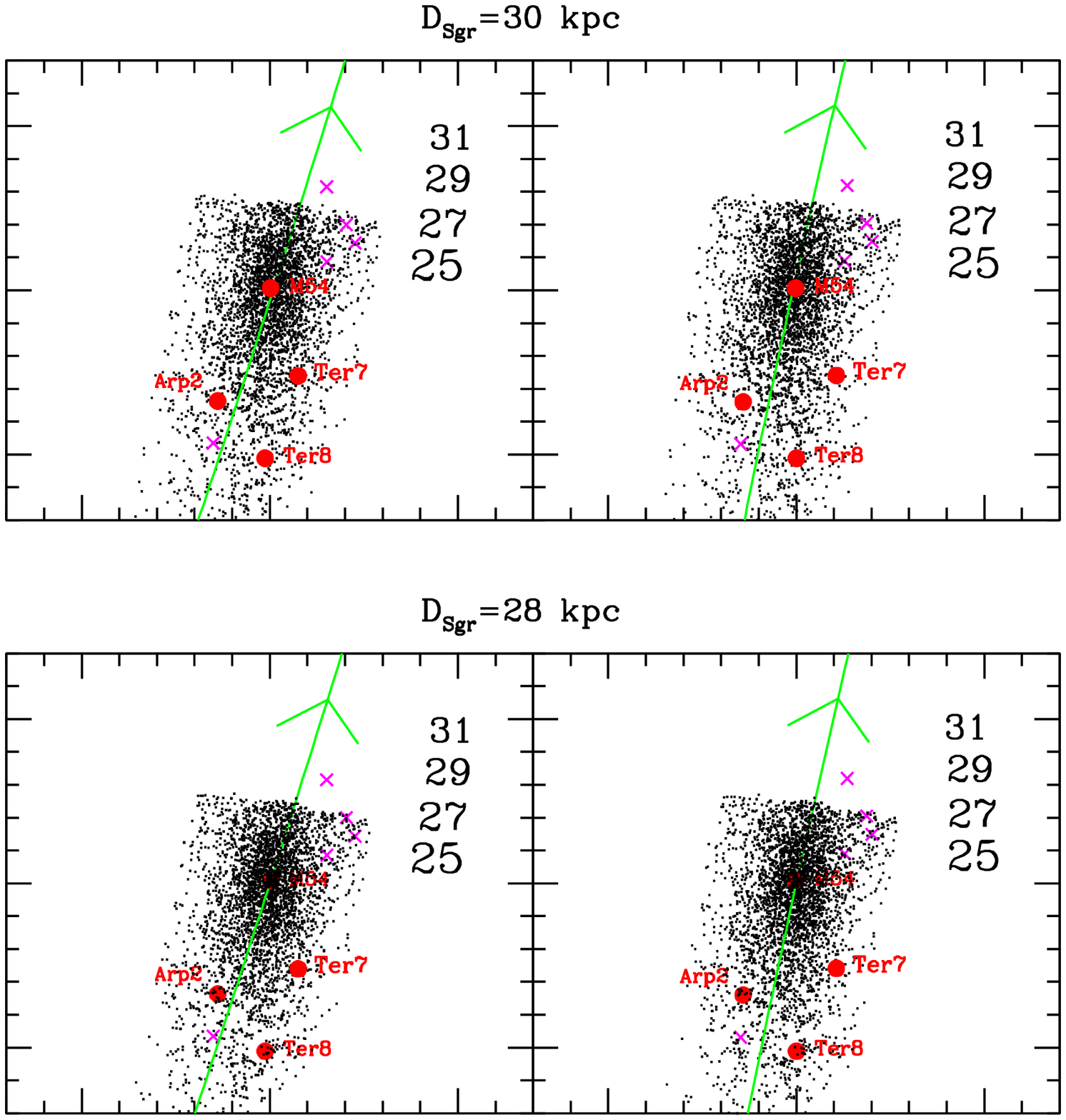}
\figcaption[f16.eps]{The Sgr dwarf core and globular cluster system in a stereoscopic representation
for a viewer at the location of the Sun with `eyes' 2.5 kpc apart along the Galactic Y-axis
(the direction of motion of the LSR).  The view approximately has Galactic longitude along the horizontal axis
(increasing from right to left) and Galactic latitude along the vertical axis (increasing from bottom to top), with a total field of
view $\sim 30^{\circ} \times 30^{\circ}$.
Blue crosses represent 
Sgr distances measured from the Sgr main sequence features; red dots are the classical Sgr core globular clusters.
Black points represent 2MASS M-giants scaled 
to a Sgr core distance of the LM10a N-body model of $d_{\rm Sgr} = 28$ kpc (lower panel) and $d_{\rm Sgr} = 30$ kpc (upper panel).
The M giant distribution at the top of the image has been cut off because of contamination by disk M giants.
The green line shows the orbit of Sgr, with the direction of motion indicated by the arrow.
The depth gauge indicates the appearance of objects at distances of 25-31 kpc.
These images are constructed in the cross-eyed stereoscopic format; the viewer can check that distances appear
correctly by verifying that the 25 kpc distance scale marker appears closer than the 31 kpc marker.
Both parallel
view and cross-eyed view versions of this image are available at http://www.astro.virginia.edu/\~srm4n/Sgr/stereo.html.
\label{f:sgr3d}}
\end{figure}
\end{center}

To aid the reader in visualizing the relative three-dimensional orientation that our data imply for Sgr's inner regions,
we have created a stereoscopic image of the Sgr system using our HST distance measures for the globular clusters and bulge cluster background
CMD features,
and the positions of the 2MASS M-giants (from Figure 4 of M03) with distances scaled to the LM10a simulation (Figure \ref{f:sgr3d}).
The top panel shows a stereoscopic image with Sgr set at a distance of 30 kpc; the bottom panel at 28 kpc.
In these images, the green line represents the orbital path of the Sgr core with an arrow indicating its direction (i.e., towards the Galactic plane).
The blue crosses show the relative distances measured for
Sgr main sequence debris stars in the background of the bulge clusters, the red dots are the classical Sgr clusters and the points are 2MASS giants.
Distances were assigned to the M giants schematically and guided by the N-body models so that, for each M giant, the 
distance corresponds to the distance of the N-body model point that lies closest in projection.
The stereoscopic image shows that M54 is either centered in the core of
the Sgr dSph (bottom panel), or projected a few kpc in front of the core (top panel) depending on the chosen distance of the Sgr core.
Our data clearly favor the longer distance -- as evidenced in 
the stereoscopic image by the blue crosses lying among the M giants in depth perception for the 30 kpc image, but behind the M giants in the 28 kpc image.
The stereoscopic image also makes evident that Terzan 7 is well in the foreground of the Sgr stars and other Sgr globular clusters no matter whether a 28 kpc or 30 kpc distance
is adopted for the Sgr center.  The M giants in the Sgr tail at the bottom of the image appear closer, following the general orientation of the Sgr
orbit.\footnote{Both parallel
view and cross-eyed view versions of this image are available at http://www.astro.virginia.edu/\~srm4n/Sgr/stereo.html. To view the cross-eyed version
presented here, the reader should slowly let their eyes cross until the images are doubled, then continue crossing their eyes until the middle images overlap,
creating a 3-D perspective.}

Further studies of other fields within the Sgr core, perhaps using other bulge clusters as foreground reference points, would provide
additional critical constraints on the emerging tidal debris.  $N$-body simulations using a greater variety of Sgr
core shapes and sizes would also allow a more robust comparison of the predicted properties of Sgr debris to our new measures.
Finally, larger area surveys, such as the VVV survey of Minniti et al. (2010), will provide additional insight into the large
and small-scale structure of this intriguing dSph satellite galaxy of the Milky Way.

\acknowledgements

Support for this work (proposal number GO-10775) was provided by NASA through a grant from the Space 
Telescope Science Institute which is operated by the Association of Universities for Research 
in Astronomy, Incorporated, under NASA contract NAS5-26555. DRL acknowledges support provided by
NASA through Hubble Fellowship grant HF-51244.01 awarded by the Space Telescope Science Institute.
MHS was supported at PSU by NASA contract NAS5-00136. SRM was supported by NSF grant AST-0807945.
This paper arose from discussion and work performed at the Aspen Center for Physics, which is 
supported by NSF and NASA funding.

\end{document}